\documentclass[aps,pra,amsmath,amssymb,floatfix,nofootinbib,notitlepage,superscriptaddress,10pt]{revtex4-2}
\pdfoutput=1
\usepackage{amsthm}
\usepackage{amsmath}
\usepackage{amssymb}
\usepackage{latexsym}
\usepackage{amsfonts}
\usepackage{bbm,dsfont}
\usepackage{graphicx,xcolor,subfigure}
\usepackage{braket}
\usepackage{hyperref}
\usepackage[normalem]{ulem}
\usepackage{verbatim}
\usepackage{bm}
\definecolor{bottle_green}{RGB}{0,106,78}
\definecolor{celadon_green}{RGB}{47,132,124}
\definecolor{emerald}{RGB}{80,220,100}
\definecolor{jade}{RGB}{0,168,107}
\hypersetup{colorlinks,
linkcolor=bottle_green,
citecolor=celadon_green,
urlcolor=emerald,
anchorcolor=jade}
\newtheorem{theorem}{Theorem}
\newtheorem{lemma}[theorem]{Lemma}
\newtheorem{definition}{Definition}
\newtheorem{proposition}[theorem]{Proposition}

\newcommand{\Tr}[2]{\mathrm{Tr}_{#1}\left[ {#2} \right]}

\newcommand{\id}{\mathds{1}}

\newcommand{\bg}[1]{\boldsymbol{#1}}
\newcommand{\abs}[1]{\left\vert#1\right\vert}

\newcommand{\mrm}[1]{\mathrm{#1}}
\newcommand{\gto}{\xrightarrow{GTO}}
\newcommand{\pos}[1]{\left[ #1 \right]^+}
\begin{document}
%%%%%%%%%%%%%%
\title{Catalytic Gaussian thermal operations}
\date{\today}
%%%%%%%%%%%%%%
\author{Benjamin Yadin}
\thanks{These authors contributed equally}
\email{benjamin.yadin@gmail.com}
\email{h.jee17@imperial.ac.uk}
\affiliation{Wolfson College, University of Oxford, Linton Road, Oxford OX2 6UD, United Kingdom}
\affiliation{Naturwissenschaftlich-Technische Fakult\"at, Universit\"at Siegen, Walter-Flex-Straße 3, 57068 Siegen, Germany}

\author{Hyejung H.\ Jee}
\thanks{These authors contributed equally}
\email{benjamin.yadin@gmail.com}
\email{h.jee17@imperial.ac.uk}
\affiliation{Department of Computing, Imperial College London, London SW7 2AZ, UK}

\author{Carlo Sparaciari}
\affiliation{Department of Computing, Imperial College London, London SW7 2AZ, UK}
\affiliation{Department of Physics and Astronomy, University College London, London WC1E 6BT, UK}

\author{Gerardo Adesso}
\affiliation{School of Mathematical Sciences, University of Nottingham, University Park, Nottingham NG7 2RD, UK}

\author{Alessio Serafini}
\affiliation{Department of Physics \& Astronomy, University College London, Gower Street, London WC1E 6BT, UK}

%%%%%%%%%%%%%%
\begin{abstract}

We examine the problem of state transformations in the framework of Gaussian thermal resource theory in the presence of catalysts. To this end, we introduce an expedient parametrisation of covariance matrices in terms of principal mode temperatures and asymmetries, and consider both weak and strong catalytic scenarios. We show that strong catalysts (where final correlations with the system are forbidden) are useless for the single mode case, in that they do not expand the set of states reachable from a given initial state through Gaussian thermal operations. We then go on to prove that weak catalysts (where final correlations with the system are allowed) are instead capable of reaching more final system states, and determine exact conditions for state transformations of a single-mode in their presence. Next, we derive necessary conditions for Gaussian thermal state transformations holding for any number of modes, for strong catalysts and approximate transformations, and for weak catalysts with and without the addition of a thermal bath. We discuss the implications of these results for devices operating with Gaussian elements.
    
\end{abstract}
%%%%%%%%%%%%%%
\maketitle
%\tableofcontents
%%%%%%%%%%%%%%

\section{Introduction}
The extension of thermodynamics outside of its original classical scope to include nano-scale, fluctuating, and quantum systems has required the development of novel conceptual and theoretical approaches~\cite{Vinjanampathy2016Quantum,Goold2016Role}.
In this regard, quantum resources theories~\cite{Chitambar2019Quantum}, growing out of quantum information, have attracted attention recently for their application to thermodynamics~\cite{Horodecki2013Fundamental}.
Indeed, the resource theory approach enables precise definitions of work and heat as well as novel insights into the role of quantum statistical quantities in small systems, such as entropy, entanglement, and coherence~\cite{Lostaglio2015Quantum}.
Moreover, it led to the appropriate mathematical formulation of the second law in quantum thermodynamics in terms of a family of free energy quantities~\cite{Brandao2015Second}.
However, these works have typically considered discrete-variable systems with the assumption that any energy-conserving dynamics are in principle possible.
Thus, they have limited applicability to continuous-variable (CV) systems, some of the most important physical platforms for testing quantum thermodynamics and for quantum technologies more generally.
These include harmonic oscillators ranging from optics to mechanical systems; in such systems, the range of available operations and interactions is often limited to Gaussian elements~\cite{Weedbrook2012Gaussian,Adesso2014Continuous,bucca}.
This argument has motivated the formulation of restricted Gaussian resource theories in a number of fields~\cite{Lami2018Gaussian,Takagi2018Convex,Giedke2002Characterization}, and thermodynamics also requires a better adapted resource theory in the Gaussian regime.

Outside of the resource theory context, the thermodynamics of Gaussian CV systems has begun to be understood in various ways, including heat engines operating with squeezed reservoirs~\cite{Rossnagel2014Nanoscale,Abah2014Efficiency,Klaers2017Squeezed,Manzano2016Entropy,Niedenzu2018Quantum}, studies of networks~\cite{Martinez2013Dynamics,Freitas2017Fundamental}, and via the notion of passivity~\cite{Friis2018Precision,Brown2016Passivity,Singh2019Quantum}.
However, to unify these approaches and find general fundamental limits will require the relevant CV resource theory to be fully developed. The first steps in this direction were made in Refs.~\cite{Serafini2020Gaussian,Narasimhachar2021Thermodynamic}, which defined the resource theory of thermodynamics in Gaussian CV systems.
Ref.~\cite{Narasimhachar2021Thermodynamic} identified a set of thermodynamical resources under passive linear interactions with a thermal environment and found some second law-like statements, while Ref.~\cite{Serafini2020Gaussian} analysed a slightly more general context with arbitrary quadratic Hamiltonians and provided a full characterisation of the possible dynamics of a single mode.

One element not yet considered in these works is the role played by catalysts.
The notion of a catalyst -- a system that can be used but must be returned back to its initial state -- is central even to classical thermodynamics, applying to any machine that operates in a cycle.
The importance of catalysts to quantum resource theories beyond thermodynamics was demonstrated famously in entanglement theory~\cite{Jonathan1999Entanglement}; in general, allowing catalysts realises more possible state transformations under the same restricted set of operations.
In discrete-variable thermodynamics, it is known that one recovers a single inequality describing the second law involving the Helmholtz free energy when looking at approximate transitions in the identical and independently distributed (i.i.d.) limit~\cite{Brandao2015Second} or we allow for correlations between system and catalyst in the final state~\cite{Muller2018Correlating}.
Other interesting phenomena with catalysts include embezzlement, which occurs if one is not careful with the error bound for approximate catalysts~\cite{Brandao2015Second,Ng2015Limits}, a catalytic-type property obeyed by coherence~\cite{Aberg2014Catalytic}, and a universality enjoyed by all catalyst states given sufficiently many copies~\cite{LipkaBartosik2021All}.

In this paper, we characterise possible state transformations for Gaussian states under Gaussian thermal operations when catalysts are allowed. Our work can be considered an extension of the initial studies~\cite{Serafini2020Gaussian,Narasimhachar2021Thermodynamic} in two ways: (i) we now allow for catalysts, and (ii) we obtain state transition conditions for the multi-mode case. As the possible state transformations under non-catalytic Gaussian thermal operations characterised in \cite{Serafini2020Gaussian, Narasimhachar2021Thermodynamic} were very limited, we ask whether catalysts can help us to perform more interesting thermodynamic tasks in the Gaussian regime. We consider two different models for catalysts. The first type is called a \emph{strong catalyst}, which must not only come back to its original state but also end up uncorrelated from the system. The second type, a \emph{weak catalyst}, allows for final correlation with the system and requires the catalyst only locally to return to its initial state. Despite the general usefulness of catalysts, we find that the state transitions we can achieve with strong catalysts are exactly the same as those without catalysts, while we can achieve more -- yet still limited -- state transformations with weak catalysts. Further, we determine the full necessary and sufficient conditions for single-mode catalytic Gaussian thermal transformations. For the multi-mode case, we introduce a new set of resource monotones with clear physical interpretation and provide state transition conditions in terms of them.

This paper is structured as follows. In Section~\ref{sec:prel}, we revise the definitions and properties of Gaussian thermal operations and introduce a useful representation of Gaussian states and operations. We first look at possible state transformations under non-catalytic Gaussian thermal operations in Section~\ref{sec:noncat} and then formally define two different types of catalysts in Section~\ref{sec:cat_def}. The full characterisation of state transformations under single-mode catalytic Gaussian thermal operations is discussed in Section~\ref{sec:single_mode}, and the state transition conditions in the multi-mode case are presented in Section~\ref{sec:multi_mode}. Finally, we discuss the physical implications of our results in Section~\ref{sec:physical_implications} and then conclude the paper in Section~\ref{sec:conclusions}.

%%%%%%%%%%%%%%
\section{Preliminaries}\label{sec:prel}
\subsection{Gaussian thermal operations (GTO)}
We consider an $n$-mode bosonic continuous-variable (CV) quantum system associated with a tensor product Hilbert space $\mathcal{H}^{\otimes n}=\bigotimes_{k=1}^{n}\mathcal{H}_k$ and the corresponding set of Hermitian quadrature operators $\{\hat{x}_i\}_i$ and $\{\hat{p}_i\}_i$. If we define a vector with these operators, $\hat{\mathbf{r}} = (\hat{x}_1,\hat{p}_1,...,\hat{x}_n,\hat{p}_n)^T$, it satisfies the bosonic commutation relations $[\hat{r}_i , \hat{r}_j]= i \Omega_{ij}$, equivalently $[\hat{\mathbf{r}},\hat{\mathbf{r}}^T]=i\Omega$ with the symplectic form matrix $\Omega=\Omega_1^{\oplus n}$ where $\Omega_1=\begin{pmatrix}0&1\\-1&0\end{pmatrix}$. The quadrature operators are related to the bosonic field operators $\{\hat{a}_i,\hat{a}_i^{\dagger}\}_i$ via $\hat{a}_i = \frac{\hat{x}_i+i\hat{p}_i}{\sqrt{2}}$.
For a generic quantum state $\rho$, we define a vector of first moments as
\begin{align}
    {\bf r} := \langle\hat{\bf r}\rangle_{\rho} = \left(\langle\hat{x}_1\rangle_{\rho}, \langle\hat{p}_1\rangle_{\rho},..., \langle \hat{x}_n\rangle_{\rho}, \langle\hat{p}_n\rangle_{\rho} \right),
\end{align}
and the covariance matrix (CM) $\sigma$ as
\begin{align}
    \sigma_{ij} := \frac{1}{2}\left\langle \left\{ \hat{x}_i-\langle\hat{x}_i\rangle_{\rho}, \hat{x}_j-\langle\hat{x}_j\rangle_{\rho} \right\}\right\rangle_{\rho},
\end{align}
where $\{\cdot,\cdot\}$ is the anti-commutator. A CM is physically realisable when it satisfies the uncertainty principle, which can be written as $\sigma+\frac{i}{2}\Omega\geq0$ \cite{Simon1994Quantum}.
Gaussian states are defined as the ground or thermal states of a second-order Hamiltonian $\hat{H}=\frac{1}{2}(\hat{\bf r}-{\bf r}_0)^TH(\hat{\bf r}-{\bf r}_0)$ with a symmetric Hamiltonian matrix $H$ and a real vector ${\bf r}_0$ \cite{bucca}. They can be fully characterised by their first moments and CMs.

Gaussian operations are all quantum operations (i.e., completely positive trace-nonincreasing maps) which map Gaussian states into Gaussian states~\cite{Adesso2014Continuous,bucca}.
Any Gaussian unitary operation admits a representation with a real $2n \times 2n$ symplectic matrix $S\in\mathrm{Sp(2n)}$, meaning that it preserves the commutations relations, $S \Omega S^T = \Omega$. Its action on a Gaussian state with first moments ${\bf r}$ and CM $\sigma$ can be described by ${\bf r}\mapsto S{\bf r}$ and $\sigma\mapsto S\sigma S^T$.
The sub-class of \textbf{passive linear unitaries} describes all operations implemented in optics with phase shifts and beam-splitters -- in other words, Gaussian unitaries that do not involve squeezing~\cite{Reck1994Experimental}.
These are represented by the set of operators that are both symplectic and orthogonal, $\mathrm{K}(n) := \operatorname{Sp}(2n) \cap \mathrm{O}(2n)$.
Since $\mathrm{K}(n)$ is isomorphic to $\mathrm{U}(n)$, any passive linear unitary can be identified with an $n \times n$ unitary matrix $U$ under which the ladder operators transform in the Heisenberg picture as $\hat{a}_i \to \sum_j U_{ij} \hat{a}_j$.

In general, a Gaussian operation can be implemented by letting a system interact with an environment in a Gaussian state under a joint Gaussian unitary and then measuring the environment by projecting onto Gaussian states~\cite{Weedbrook2012Gaussian,bucca}.
For the thermodynamical context, we consider the intersection of these with the \textbf{thermal operations} which define the relevant resource theory for discrete-variable systems~\cite{Horodecki2013Fundamental}.
Thermal operations put minimal restrictions on processes in which energetic and entropic resources can be tracked explicitly: a system is permitted to interact via any global energy-conserving unitary with an environment (of any dimension and Hamiltonian) starting in a thermal state at some fixed background temperature\footnote{Note that measurements are not permitted for free.}.

\textbf{Gaussian thermal operations} (GTOs) are a sub-class of Gaussian operations which can be realised by energy-preserving interaction between a system and a thermal bath. More specifically, for a given second-order system Hamiltonian $\hat{H}_\mrm{S}$ and an inverse temperature $\beta = 1/k_BT$, GTOs are defined as operations obtained by (i) preparing a thermal state with arbitrary second-order Hamiltonian $\hat{H}_\mrm{B}$, i.e., $e^{-\beta\hat{H}_\mrm{B}}/\Tr{}{e^{-\beta\hat{H}_\mrm{B}}}$, and (ii) applying an energy-preserving Gaussian unitary $\hat{U}_\mrm{SB}=e^{-i\hat{H}_\mrm{SB}t}$ such that $[\hat{H}_\mrm{SB}, \hat{H}_\mrm{S}+\hat{H}_\mrm{B}]=0$.
The following is a representation of GTOs acting on covariance matrices:

\begin{theorem}\label{thm:generalGTO}
\cite{Serafini2020Gaussian} Let $\hat{H}_\mrm{S}=\frac{1}{2}\hat{\mathbf{r}}^TH_\mrm{S}\hat{\mathbf{r}}$ be a system Hamiltonian with normal form $S^{-1}H_\mrm{S}(S^T)^{-1}=\oplus_l\omega_l\id_{2n_l}$ where $n_l$ is the mode degeneracy of an eigenfrequency $\omega_l$. For a given system CM $\sigma$, the transformation under generic GTOs at background inverse temperature $\beta$ can be characterised by
\begin{align}\label{eq:generalGTO}
    \sigma \mapsto S\left(\oplus_l W_l \circ\Phi_l\circ Z_l \left(S^{-1}\left(\sigma\right)\right)\right),
\end{align}
where $W_l$ and $Z_l$ are passive linear unitaries acting only on the modes with eigenfrequency $\omega_l$, and $\Phi_l$ are CP maps describing thermalisation as a result of interacting with thermal baths, as follows:
\begin{align}
    \Phi_l(\sigma_l) = X_l \sigma_l X_l^{T} + Y_l ,
\end{align}
where $\sigma_l$ is the CM of the $l$-th degenerate sector, $X_l = \bigoplus_{k=1}^{n_l}\cos\theta_{lk}\id_2$ and $Y_l = \bigoplus_{k=1}^{n_l}\frac{{\rm e}^{\beta\omega_l}+1}{{\rm e}^{\beta\omega_l}-1}\sin^2\theta_{lk}\id_2$, for some $\theta_{lk}\in {\mathbbm R}$. 
\end{theorem}

\noindent Note that in Eq.~\eqref{eq:generalGTO}, apart from the symplectic transformation $S$ which brings the system Hamiltonian into the normal form, modes with different eigenfrequencies transform independently -- GTOs cannot make modes with different eigenfrequencies interact with each other. The characterisation in Theorem~\ref{thm:generalGTO} says that, once the Hamiltonian is transformed to its normal form, a GTO can be implemented in three steps: (i) adding thermal bath ancillae with same eigenfrequency and same number of modes to each different eigenfrequency sector of the system; (ii) applying passive linear unitaries to each eigenfrequency sector separately; (iii) tracing out the bath modes.

Since non-degenerate modes do not interact, in this work we only consider the interesting case of degenerate modes -- so we assume that all modes have the same eigenfrequency.
Moreover, we assume a mode basis has already been chosen such that the Hamiltonian is in its normal form $H_\mrm{S} = \omega \id_{2n}$; equivalently $\hat{H}_\mrm{S} = \omega (\hat{N}_\mrm{S} + 1/2)$, where $\hat{N}_\mrm{S}$ is the total number operator.
Under these conditions GTOs effectively reduce to only the three types of operations described above, which were introduced as bosonic linear thermal operations in Ref.~\cite{Narasimhachar2021Thermodynamic}.

\subsection{State representation: principal mode temperatures and principal mode asymmetries}\label{subsec:PMTemperatures_PMAsymmetries}
A generic Gaussian state can be represented by its vector of first moments $\mathbf{r}$ and its CM $\sigma$. 
Here, we consider the case of vanishing first moments, $\mathbf{r}=0$, and concentrate only on the thermodynamical properties of the CM.
We employ a decomposition of the CM, first introduced in Ref.~\cite{Simon1994Quantum}, that provides both a convenient mathematical description for analysing transformations under GTOs and a set of quantities with interesting physical interpretations.
A $2n\times2n$ covariance matrix can be decomposed into two $n\times n$ matrices:
\begin{align}
    M_{ij}:=\left\langle\hat{a}^{\dagger}_j\hat{a}_i\right\rangle - \delta_{ij}\nu\,, \quad A_{ij}:=\left\langle \hat{a
    }_j \hat{a}_i \right\rangle\,,
\end{align}
where $\nu=\frac{e^{\beta\omega}+1}{e^{\beta\omega}-1}$ is the variance of any quadrature in a thermal state at background inverse temperature $\beta$. They are both complex matrices; $M$ is Hermitian, and $A$ is symmetric. $M$ is sometimes known as the single-particle density matrix, up to a constant shift. In terms of these two matrices, we can recover the CM $\sigma$ via
\begin{align}
    \left\langle \hat{x}_i\hat{x}_j \right\rangle &= \Re[M_{ij}+A_{ij}] + \delta_{ij}\nu\,, \\
    \left\langle \hat{p}_i\hat{p}_j \right\rangle &= \Re[M_{ij}-A_{ij}] + \delta_{ij}\nu\,, \\
    \frac{1}{2}\left\langle \{\hat{x}_i,\hat{p}_j\} \right\rangle &= \Im[A_{\ij}-M_{ij}].
\end{align}
This representation is convenient since a thermal state at the background temperature is characterised by $M=A=0$.
Furthermore, these matrices evolve simply under the passive linear unitaries on which GTOs are based.
Under such an operation described by a unitary matrix $U$, the matrices $M$ and $A$ transform as
\begin{alignat}{2}
    & M_{ij} \mapsto \sum_{k,l}U_{ik}M_{kl}U^*_{jl}\,, \quad && M \mapsto UMU^{\dagger}\\
    & A_{ij} \mapsto \sum_{k,l}U_{ik}A_{kl}U_{jl}\,, \quad && A \mapsto UAU^T.
\end{alignat}
It is therefore possible to diagonalise $M$ with an appropriate choice of $U$.\footnote{Note that diagonalisation of the CM with passive linear unitaries is not in general possible, since $\text{K}(n)$ is strictly smaller than the orthogonal group $\text{O}(2n)$.} $A$ can also be diagonalised with a generally different unitary $V$.
An advantage of using $M$ and $A$ instead of $\sigma$ is that we can identify some quantities that are invariant under passive linear unitaries and thus equal for states that are thermodynamically equivalent due to being unitarily interconvertible under GTOs.

In a basis where $M$ is diagonal, we have $M=\text{diag}(\mu_1,...,\mu_n)$. $A$ is not in general diagonal in this basis, but we are free to make arbitrary phase rotations of the form $U_{ij}\rightarrow U_{ij}e^{i\phi_j}$ while leaving $M$ unchanged, under which $A_{ij}\rightarrow e^{i(\phi_i+\phi_j)}A_{ij}$. We can thus always leverage this freedom to make the diagonals real and non-negative, i.e., $A_{ii}\geq0$. Ordering the quadratures as $(\hat{x}_1,\hat{p}_1,\hat{x}_2,\hat{p}_2,...)$, the resulting form of the covariance matrix is
\begin{equation} \label{eq:CM_in_normal_form}
	\sigma = \left( \begin{array}{cc|cc|cc}
		\mu_1 + A_{11} + \nu	& 0	 			& \Re[A_{12}]	& \Im[A_{12}]		& \cdots	& \cdots \\
		0	& \mu_1 -A_{11} + \nu				& \Im[A_{12}]	& -\Re[A_{12}]		& \cdots & \cdots \\ \hline
		\Re[A_{12}]	& \Im[A_{12}]	& \mu_2 + A_{22} + \nu	& 0						& \cdots & \cdots \\
		\Im[A_{12}]	& -\Re[A_{12}]	& 0		& \mu_2 - A_{22} +\nu					& \cdots & \cdots \\ \hline
		\vdots 		& \vdots						& \vdots 	& \vdots	& \ddots & \ddots \\
		\vdots 		& \vdots						& \vdots 	& \vdots	& \ddots & \ddots \\		
	\end{array} \right).
\end{equation}
Aside from the eigenvalues $\{\mu_i\}_i$ of $M$, another set of invariant quantities under passive linear operations are the singular values $\alpha_i\geq0$ of $A$, obtained by diagonalising $A$ with the unitary $V$ by congruence~\cite[Chapter 4]{Horn1985Matrix}. In this paper, we characterise states in terms of the parameters $\bm{\mu}=(\mu_1,...,\mu_n)$ and $\bm{\alpha}=(\alpha_1,...,\alpha_n)$. Without loss of generality, we assume that they are arranged in descending order. Individually, $\bm{\mu}$ and $\bm{\alpha}$ determine $M$ and $A$ up to passive linear rotations\;--\;but it should be noted that they do not generally provide a full characterisation of covariance matrices. The problem is that $M$ and $A$ are not in general simultaneously diagonalisable. However, when they are, these parameters are sufficient to determine $\sigma$ up to a passive linear rotation:
\begin{definition}
A covariance matrix $\sigma$ is called \textbf{decouplable} if there exists a passive unitary that makes all the modes uncorrelated. In this case, the normal form in Eq.~\eqref{eq:CM_in_normal_form} is fully diagonal and $A_{ii}=\alpha_i$ for all $i$.
\end{definition}

\begin{figure}[h]
    \includegraphics[width=0.4\textwidth]{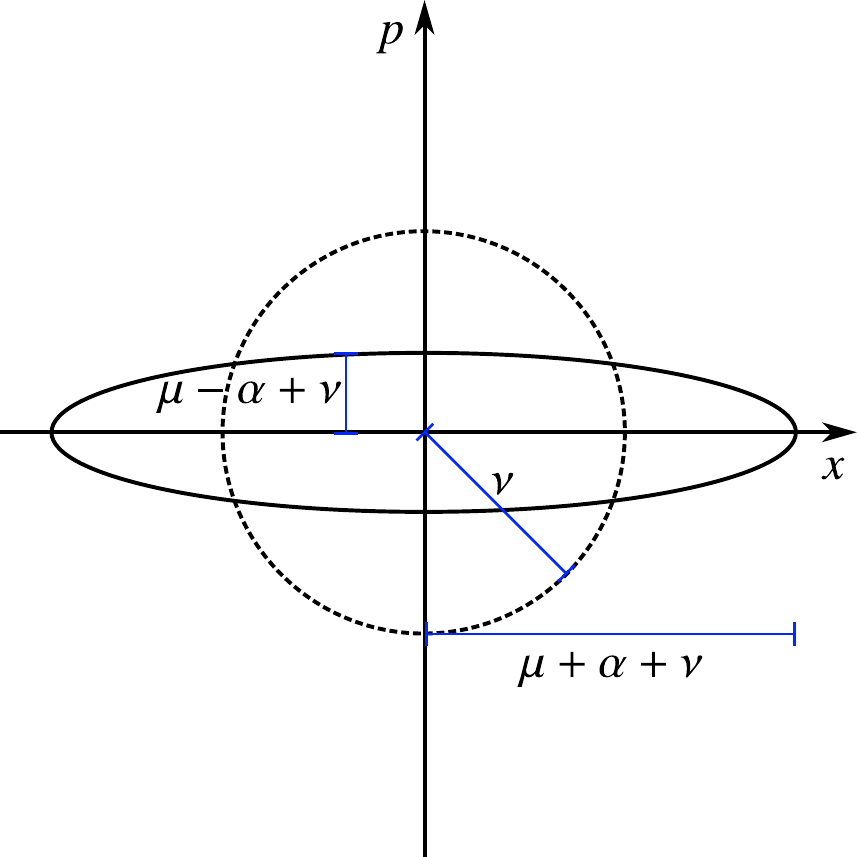}
    \caption{Illustration of the parameters used to characterise covariance matrices (shown for a single mode). We use a common visual representation of Gaussian states where an ellipse indicates the size of fluctuations in every quadrature direction. Note that the quantities indicated are the squares of the lengths. Dashed circle: thermal state at the background temperature $T$, with variance $\nu$ in all quadrature directions. Solid ellipse: arbitrary Gaussian state (with vanishing first moments). $\mu$ is the principal mode temperature and $\alpha$ is the principal mode asymmetry.}
    \label{fig:parameters}
\end{figure}

The parameters $\bm{\mu}$ and $\bm{\alpha}$ have interesting physical interpretations (see Fig.~\ref{fig:parameters} for an illustration). The quantities $\{\mu_i+\nu\}_i$ were named \textbf{principal mode temperatures} in \cite{Narasimhachar2021Thermodynamic} and described as the most extreme effective temperatures that can be found in any mode decomposition of a state. The mean energy of the $i$-th mode is a function of $\left\langle\hat{x}_i^2+\hat{p}_i^2\right\rangle/2=M_{ii}+\nu$ in the case of vanishing first moments; phase-averaging the mode results in a thermal state with the same mean energy.
Note that this quantity is directly related to an effective temperature bounding the efficiency of a proposed heat engine using general Gaussian reservoirs~\cite{Rossnagel2014Nanoscale,Abah2014Efficiency}.
Given the ordering $\mu_1 \geq \mu_2 \geq \dots$, $\mu_1$ is the greatest effective temperature found in any mode decomposition.
$\mu_2$ is the next greatest value found in any mode orthogonal to this, and so on.

The $\{\alpha_i\}_i$ instead describe the rotational asymmetry of modes in phase space, since $\left\langle\hat{x}_i^2-\hat{p}_i^2\right\rangle/2=\Re[A_{ii}]=A_{ii}$ using the aforementioned phase freedom. We name them \textbf{principal mode asymmetries} in this paper.
Again ordering $\alpha_1 \geq \alpha_2 \geq \dots$, we see that $\alpha_1$ describes the mode with the greatest asymmetry, $\alpha_2$ the next greatest amongst modes orthogonal to that, and so on.
These are somewhat similar to squeezing, except that $\alpha>0$ does not necessarily imply a sub-shot-noise quadrature variance.

%%%%%%%%%%%%%%

\section{Gaussian thermal operations without catalysts} \label{sec:noncat}
In general, we would like to understand when one state $\rho$ of a system can be transformed into another state $\rho'$ via a GTO in the presence of catalysts when all modes are degenerate. In this section, we revise state transformations under normal (non-catalytic) GTOs, focussing on the principal mode temperatures and asymmetries.

When there exists a GTO which maps a state $\rho$ of a system to another state $\rho'$, we write $\rho \gto \rho'$.
For Gaussian states with vanishing first moments, this is equivalent to the conversion of one CM into another, denoted $\sigma \gto \sigma'$.
Alternatively, a simpler question can be asked using our decomposition of the CM: does a GTO exist that maps a state with matrix $M$ into one with matrix $M'$, not concerned with what happens to the matrix $A$?
When this is possible, we write $M \gto M'$; likewise, the notation $A \gto A'$ implies the matrix $A$ can be transformed into $A'$ through GTOs.

The following result from Ref.~\cite{Serafini2020Gaussian} (restated in terms of the parameters used here) answers when one CM of a single mode can be converted into another under GTO (notice that the matrices $A$ and $M$, reducing to single numbers, necessarily commute for a single mode):
\begin{proposition}\label{pro:single_mode_GTO}
    \cite{Serafini2020Gaussian} Let $\sigma,\, \sigma'$ be covariance matrices of a single mode. Then $\sigma \gto \sigma'$ if and only if there exists $p \in [0,1]$ such that
    \begin{align}\label{eq:single_mode_GTO_condition_in_mualpha}
        \mu' = p \mu, \quad \alpha' = p \alpha.
    \end{align}
\end{proposition}
\noindent One can state this condition alternatively as the existence of $p \in [0,1]$ and $S \in \mrm{K}(1)$ such that\
\begin{equation}\label{eq:single_mode_GTO_condition_in_CM}
    \sigma' = p\, S \sigma S^T + (1-p) \, \nu \id_2\,.
\end{equation}
In words, any single-mode GTO is equivalent to applying a phase rotation and mixing with thermal noise.

For multiple modes, a set of necessary conditions in terms of principal mode temperatures were found in Ref.~\cite{Narasimhachar2021Thermodynamic}.
We prove them again here with a different method that we also generalise to the principal mode asymmetries.
We first need to introduce some useful notation.
The parameters $\bm{\mu}$ can be divided into \textbf{super- and sub-thermal} (or ``hot" and ``cold") parts, namely the positive and negative values: $\bm{\mu} = (\mu^+_1, \dots, \mu^+_{n_+}, 0, \dots, 0, -\mu^-_{n_-}, \dots, -\mu^-_1)$, where
\begin{align}
    \mu^+_1 & \geq \mu^+_2 \geq \dots \geq \mu^+_{n_+} > 0, \nonumber \\
    \mu^-_1 & \geq \mu^-_2 \geq \dots \geq \mu^-_{n_-} > 0.
\end{align}
It will sometimes be convenient to extend these lists with zeroes, so that by convention $\mu^\pm_i = 0$ for $i > n_\pm$.

The result presented here involves a kind of ordering between such vectors composed of non-increasing elements: for $\bm{x}$ and $\bm{y}$, we write $\bm{x}\leq\bm{y}$ when $x_i\leq y_i$ for all $i$.

\begin{theorem} \label{thm:no_catalyst}
    \begin{enumerate}
        \item (Necessity proved in \cite{Narasimhachar2021Thermodynamic}) For $M$ with eigenvalue vector $\bm{\mu}$ and $M'$ with $\bm{\mu'}$, $M\xrightarrow{GTO}M'$ if and only if $\bm{\mu'}^+\leq\bm{\mu}^+$ and $\bm{\mu'}^-\leq\bm{\mu}^-$ \\
        
        \item\ (New here) For $A$ with singular-value vector $\bm{\alpha}$ and $A'$ with $\bm{\alpha'}$, $A\xrightarrow{GTO}A'$ if and only if $\bm{\alpha'}\leq\bm{\alpha}$.
    \end{enumerate}
\end{theorem}
See the proof in Appendix~\ref{app:no_catalyst}.
The main idea is that the matrices transform as $M' = P M P^\dagger,\, A' = P A P^T$, where $P$ is a submatrix of the unitary describing the coupling to the bath.
Since $P$ is a contraction, it has a corresponding contractive effect on the eigenvalues and singular values.
The sufficiency of the conditions results from finding a type of elementary GTO which applies what we call an \textbf{L-transform} on the parameters (where ``L" stands for ``lossy"), which is able to perform the required transformations applied to each mode independently.
An L-transform is achieved by simply mixing a single system mode with a thermal mode at a beam splitter of arbitrary reflectivity, and scales the parameters $\mu,\, \alpha$ towards zero.

In essence, this result says that the principal mode temperatures and asymmetries all converge individually towards the thermal values, i.e., zero. While the inequalities are necessary and sufficient for GTO transformation between $M$ or $A$ matrices independently, we are not able to say anything general about whether both sets of conditions can be \emph{simultaneously} satisfied. Thus, the inequalities are only \emph{necessary conditions} for transformation of the covariance matrix, but not sufficient.

\section{Definitions of catalytic Gaussian thermal operations} \label{sec:cat_def}
In this paper, we are interested in state transformations under catalytic GTOs. We define two different cases of catalytic GTO transformation.
\begin{definition}
\begin{enumerate}
    \item \textbf{(Strong catalysts)} We say that a strong catalytic GTO transformation from $\rho_\mrm{S}$ to $\rho'_\mrm{S}$ is possible if there exists a state $\rho_\mrm{C}$ such that
    \begin{align}
        \rho_\mrm{S}\otimes\rho_\mrm{C}\xrightarrow{GTO}\rho'_\mrm{S}\otimes\rho_\mrm{C}. 
    \end{align}
    The same terminology applies to transformations at the level of the matrices $\sigma, M$ and $A$\;---\;for example, ${M_S\oplus M_C\xrightarrow{GTO}M'_\mrm{S}\oplus M_\mrm{C}}$.
    \item \textbf{(Weak catalysts)} We say that a weak catalytic GTO transformation from $\rho_\mrm{S}$ to $\rho_\mrm{S}'$ is possible if there exist a state $\rho_\mrm{C}$ and a final state $\rho_\mrm{SC}'$ such that
    \begin{align}
        \rho_\mrm{S}\otimes\rho_\mrm{C} \xrightarrow{GTO}\rho_\mrm{SC}' \quad\text{ with }\rho_\mrm{C}'=\Tr{\mrm{S}}{\rho_\mrm{SC}'}=\rho_\mrm{C}.
    \end{align}
    The same terminology applies to transformations at the level of the matrices $\sigma, M$ and $A$\;---\; for example, $A_\mrm{S}\oplus A_\mrm{C}\xrightarrow{GTO}A_\mrm{SC}'$ with the catalyst sub-block $A_\mrm{C}'=A_\mrm{C}$.
\end{enumerate}
\end{definition}
\noindent A strong catalyst must be not only returned to its original state but also uncorrelated from the system at the end.
This agrees with the most common definition that has been used for instance in entanglement theory~\cite{Jonathan1999Entanglement}.
For a weak catalyst, arbitrary correlations between the system and the catalyst in the final state are allowed.
This weakening of the constraint has been studied recently and shown to result in sensible versions of the second law~\cite{Muller2018Correlating,Boes2019VonNeumann}.

A natural question is whether such catalytic GTO transformations allow for more state transformations to be performed on a system. For the rest of the paper, we provide necessary or sufficient conditions for these two different catalytic GTO transformations in terms of $\bm{\mu}$ and $\bm{\alpha}$ and compare them to the case where no catalyst is used.

%%%%%%%%%%%%%%
\section{Single mode with catalyst: full characterisation}\label{sec:single_mode}
We first consider the simplest case of catalytic GTO transformations, where both the system and the catalyst are composed by a single mode. This single-mode case is a special case of our formulation\;--\;since the initial state $\rho_\mrm{S}\otimes\rho_\mrm{C}$ is naturally decouplable, $\bm{\mu}=(\mu_\mrm{S},\mu_\mrm{C})$ and $\bm{\alpha}=(\alpha_\mrm{S},\alpha_\mrm{C})$ fully characterise the initial state of the system and the catalyst. 
% \hj{[Do they fully characterise the final state in the weak catalyst case as well?]} 
Indeed, the initial $M_\mrm{SC}$ and $A_\mrm{SC}$ describing the system and the catalyst can be written as 
\begin{align}
M_\mrm{SC}=
\begin{pmatrix}
\mu_\mrm{S} & 0 \\
0 & \mu_\mrm{C} 
\end{pmatrix}, \qquad
A_\mrm{SC}=
\begin{pmatrix}
\alpha_\mrm{S} & 0 \\
0 & \alpha_\mrm{C}
\end{pmatrix}.
\end{align}
Since we can perform any GTO transformation using as many bath modes as the total modes of the system and the catalyst~\cite{Serafini2020Gaussian}, we add a two-mode thermal bath at inverse temperature $\beta$. Then the joint initial state of the system, the catalyst, and the thermal bath is described by the following $M$ and $A$ matrices:
\begin{align}
M = 
\begin{pmatrix}
\mu_\mrm{S} & 0 & 0 & 0 \\
0 & \mu_\mrm{C} & 0 & 0 \\
0 & 0 & 0 & 0 \\
0 & 0 & 0 & 0
\end{pmatrix}
= M_\mrm{SC}\oplus \boldsymbol{0}_\mrm{B}\,\qquad
A =
\begin{pmatrix}
\alpha_\mrm{S} & 0 & 0 & 0 \\
0 & \alpha_\mrm{C} & 0 & 0 \\
0 & 0 & 0 & 0 \\
0 & 0 & 0 & 0
\end{pmatrix}
=A_\mrm{SC}\oplus \boldsymbol{0}_\mrm{B}\,,
\end{align}
where $\boldsymbol{0}_\mrm{B}$ denotes here the $2\times2$ null matrix. As discussed in Section~\ref{sec:prel}, any GTO transformation can be obtained by applying a passive linear unitary to the whole state and tracing out the bath modes, and any passive linear unitary corresponds to a unitary $U$ acting on $M$ and $A$ matrices. Let us divide the interacting passive operation $U$ into four $2\times2$ blocks,
\begin{align}
U = 
\begin{pmatrix}
P & Q \\
R & S
\end{pmatrix}.
\end{align}
The submatrix $P$ satisfies $P^{\dagger}P \leq \id$ and $PP^{\dagger}\leq\id$ (denoting the usual matrix inequality).
With this notation, the evolution of the system and the catalyst is only affected by the sub-matrix $P$:
\begin{align}\label{eq:MandA_transformation_via_P}
    M_\mrm{SC}'=PM_\mrm{SC}P^{\dagger}, \qquad A_\mrm{SC}'=PA_\mrm{SC}P^T\,.
\end{align}
The main question is: for a given description of the initial system, i.e., $\mu_\mrm{S}$ and $\alpha_\mrm{S}$, which final parameters $\mu_\mrm{S}'$ and $\alpha_\mrm{S}'$ can be reached via catalytic GTO transformations?

\begin{figure}[t]
    \centering
    \subfigure[][]{
    \label{fig:single_mode_mu_alpha}
    \includegraphics[width=0.5\linewidth]{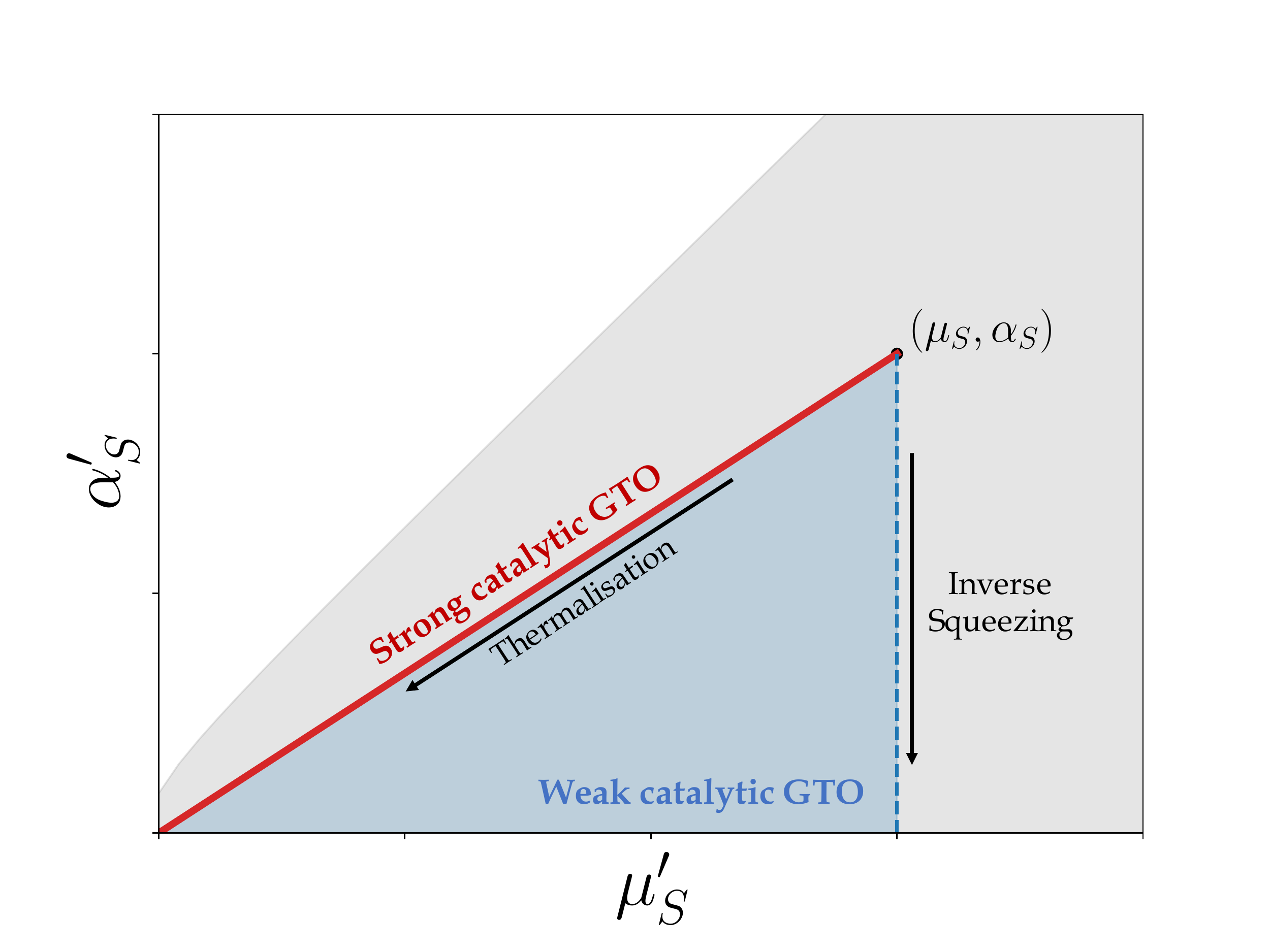}
    }
    \subfigure[][]{
    \label{fig:single_mode_CM}
    \includegraphics[width=0.45\linewidth]{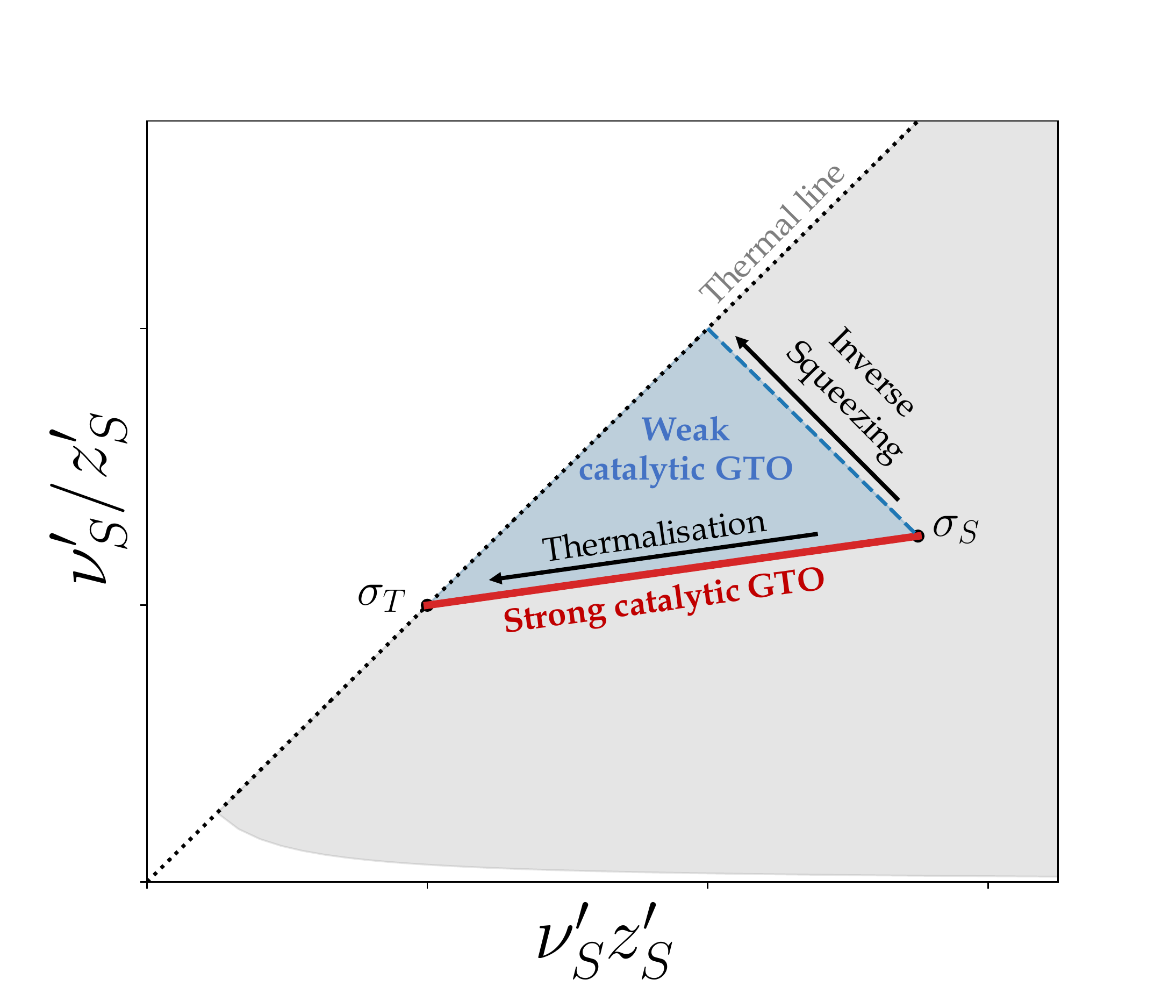}
    }
    \caption{Visualisation of possible state transformations under single-mode strong or weak catalytic GTOs, when the initial state of the system is described by the parameters $\mu_\mrm{S},\alpha_\mrm{S}$ or the CM $\sigma_\mrm{S}$. The grey regions represent physical regions in both figures.
    \subref{fig:single_mode_mu_alpha} In the state space described by the principal mode temperature $\mu$ and asymmetry $\alpha$. Each point in the physical region (grey) fully describes a covariance matrix of the single-mode system up to passive linear rotations, and the origin represents the thermal bath. The red line describes all final states achievable by strong catalytic GTOs; it is same as the non-catalytic single-mode case (thermalisation). The blue region represents all possible final states with weak catalytic GTOs. In this case, we can additionally perform inverse squeezing to the system by coupling it to the catalyst. 
    \subref{fig:single_mode_CM} In the state space described by the eigenvalues of the CM of the system. $\sigma_T$ denotes the CM of the thermal bath. Again, the red line represents all possible state transformations with strong catalytic GTOs, and the blue region describes the same with weak catalytic GTOs. Note that the red line is the same as what we could achieve with non-catalytic single-mode GTOs discussed in Proposition~\ref{pro:single_mode_GTO}.
    }
    \label{fig:single_mode}
\end{figure}

We first find that a strong catalyst does not enlarge the range of possible transformations on a single mode:
\begin{theorem}\label{thm:single_mode_strong_cat_GTO}
\textbf{(Single-mode strong catalytic GTO)} Let $\sigma$ and $\sigma'$ be covariance matrices of a single-mode bosonic Gaussian state. The state transition condition between $\sigma$ and $\sigma'$ under single-mode strong catalytic GTOs is exactly the same as the one under non-catalytic GTOs described in Eq.~\eqref{eq:single_mode_GTO_condition_in_CM}. In terms of principal mode temperatures and asymmetries, the condition is same as Eq.~\eqref{eq:single_mode_GTO_condition_in_mualpha}.

\end{theorem}
See the proof in Appendix~\ref{app:single_mode_strong_cat_GTO}.

On the other hand, with a weak catalyst, we show that a greater range of transformations becomes possible:
\begin{theorem} \label{thm:single_mode_weak_cat_GTO}
\textbf{(Single-mode weak catalytic GTO)} Consider a single-mode bosonic Gaussian system described by the parameters $\mu_\mrm{S}$ and $\alpha_\mrm{S}$. It can be mapped to the final state described by $\mu_\mrm{S}'$ and $\alpha_\mrm{S}'$ via single-mode weak catalytic GTOs if and only if there exist $p,q\in[0,1]$ such that $p\geq q$, and
\begin{align}\label{eq:single_mode_weak_catGTO}
    \mu_{\text{S}}'=p\mu_{\text{S}}, \quad \alpha_{\text{S}}'=q\alpha_{\text{S}}.
\end{align}

\end{theorem}
See the proof in Appendix~\ref{app:single_mode_weak_cat_GTO}.

These results are illustrated in Fig.~\ref{fig:single_mode_mu_alpha}.
With a strong catalyst, the only possible kind of transformation in $\mu$-$\alpha$ parameter space is to move along the line towards the origin (representing the thermal state). Physically, this implies that, apart from the symplectic transformation $S$ to the normal form, all possible state transition under single-mode strong catalytic GTOs can be described as thermalisation towards the bath mode, which is same as the case of non-catalytic single-mode GTOs.
By contrast, with a weak catalyst, a triangular region (the blue region) becomes accessible.
This requires not only $\mu$ and $\alpha$ to be non-increasing, but also the ratio $\frac{\alpha}{\mu}$.
This latter condition may be interpreted as the ``aspect ratio" of the Gaussian distribution in phase space becoming less extreme.
The same region is also redrawn in terms of the eigenvalues of the CM in Fig.~\ref{fig:single_mode_CM}.

Any point in the triangular region can be reached by concatenating two simple processes: a reduction in $\alpha$ with fixed $\mu$, followed by a non-catalytic thermalisation which scales both parameters similarly towards zero.
The first of these involves a catalyst but no bath.
In order to map $(\mu_\mrm{S},\alpha_\mrm{S}) \to (\mu_\mrm{S},\bar{\alpha}_\mrm{S})$, we take $\mu_\mrm{C} = \mu_\mrm{S}$ and $\alpha_\mrm{C} = \frac{\alpha_\mrm{S}-\bar{\alpha}_\mrm{S}}{2}$, and need the following unitary operation between these two subsystems:
     \begin{align}
            U =\begin{pmatrix}
            \sqrt{a} & i\sqrt{1-a} \\
            \sqrt{1-a} & -i\sqrt{a}
            \end{pmatrix}\,,
    \end{align}
where $a=\frac{\alpha_\mrm{S}+\bar{\alpha}_\mrm{S}}{3\alpha_\mrm{S}-\bar{\alpha}_\mrm{S}}\in[\frac{1}{3},1]$, which can be performed with a suitable beam-splitter and phase shift.

Note that, in the single-mode case, the parameters $\mu_\mrm{S}$ and $\alpha_\mrm{S}$ fully characterise the system state up to passive linear rotations. Thus, the conditions stated in Theorem~\ref{thm:single_mode_strong_cat_GTO} and \ref{thm:single_mode_weak_cat_GTO} are \emph{necessary and sufficient conditions} for state transformations under single-mode catalytic GTOs. 

%%%%%%%%%%%%%%
\section{Multiple modes with catalyst}\label{sec:multi_mode}
Here, we explore state transformations under GTOs involving systems and catalysts of arbitrarily many modes.
Given the difficulty of completely characterising the equivalence of covariance matrices under the set of passive linear unitaries, the above analysis solving the single-mode case does not generalise to multiple modes.
Instead, we determine the monotonicity properties of the quantities $\bm{\mu},\, \bm{\alpha}$ with respect to the different types of catalysts.

%%%%%%%%%%%%%%%%%%%%%%%%%%%%%%%%%%%%%%%%
\subsection{Approximate transformations with strong catalysts}

For a single mode, it was found above that a strong catalyst permits no more state transformations than are possible without a catalyst.
One might wonder whether this is due to the strict requirement that the catalyst be returned exactly to its initial state.
To probe this, we relax the condition to allow for an error with respect to the trace distance between covariance matrices.
In part, this choice of error measure is for mathematical convenience, however it is also in keeping with the idea of quadratures representing all the accessible quantities in the Gaussian context (aside from first moments).
In this context, it may be more experimentally relevant to distinguish the correlation properties of quadratures than to optimally distinguish states according to some measure such as fidelity or trace distance between states.
\begin{definition}
    \textbf{(Approximate transformations)}
    We write $M \underset{\delta}{\gto} M'$ if there exists $\tilde{M}$ such that $M \gto \tilde{M}$ and $\|M' - \tilde{M}\|_1 \leq \delta$, where $\|\cdot \|_1$ is the trace norm.
    An approximate strong catalytic transformation then takes the form
    \begin{equation}
        M_\mrm{S} \oplus M_\mrm{C} \underset{\delta}{\gto} M'_\mrm{S} \oplus M_\mrm{C}.
    \end{equation}
\end{definition}
Note that $\delta$-closeness in terms of the matrix $M$ implies $\delta$-closeness in terms of the eigenvalues, as the Wielandt-Hoffman inequality~\cite[Corollary 7.4.9]{Horn1985Matrix} for the trace norm says
\begin{equation}
    \sum_{i=1}^n \abs{\mu'_i - \mu_i} \leq \|M' - M\|_1,
\end{equation}
where as usual we assume non-increasing ordering.
The result for strong catalysts is the following:
\begin{theorem} \label{thm:multi_strong_cat}
    There exists $M_\mrm{C}$ such that $M \oplus M_\mrm{C} \underset{\delta}{\gto} M' \oplus M_\mrm{C}$ if and only if
    \begin{equation} \label{eqn:strong_cat_approx_m}
        \sum_i \pos{{\mu'}^+_i - \mu^+_i} + \sum_i \pos{{\mu'}^-_i - \mu^-_i} \leq \delta.
    \end{equation}
    Similarly, there exists $A_\mrm{C}$ such that $A \oplus A_\mrm{C} \underset{\delta}{\gto} A' \oplus A_\mrm{C}$ if and only if
    \begin{equation} \label{eqn:strong_cat_approx_a}
        \sum_i \pos{\alpha'_i - \alpha_i} \leq \delta.
    \end{equation}
    Here, $\pos{x} := \max \{x,\,0\}$ denotes the positive part of a real number.
\end{theorem}
See the proof in Appendix~\ref{app:multi_strong_cat}. The operations sufficient to perform the transformations when the inequalities are satisfied are as in the non-catalytic case of Theorem~\ref{thm:no_catalyst}.

Note that setting $\delta=0$ recovers the same transformation laws as the non-catalytic case in Theorem~\ref{thm:no_catalyst}.
Otherwise, the left-hand sides of Eq.~\eqref{eqn:strong_cat_approx_m} and Eq.~\eqref{eqn:strong_cat_approx_a} quantify the total amounts by which the laws are violated -- and these totals are bounded by the error $\delta$.
Also note that there is no possibility for embezzlement here -- such a phenomenon occurs, for instance, in entanglement theory~\cite{VanDam2003Universal}, whereby allowing for a small error in returning the catalyst state can permit arbitrary transformations on the system.
Rather, the result that a strong catalyst cannot enable any additional transformations is stable with respect to errors in the catalyst.\\

%%%%%%%%%%%%%%%%%%%%%%%%%%%%%%%%%%%%%%%%%%%
\subsection{Necessary majorisation conditions for weak catalysts}

For a weak catalyst, instead of the ordering relation denoted by the inequality symbol ($\leq$), we instead find that majorisation ($\prec$) and weak majorisation ($\prec_w$) become relevant.
Given two non-increasingly ordered vectors $\bm{x}$ and $\bm{y}$ of length $n$, we say that $\bm{y}$ \textbf{weakly majorises} $\bm{x}$, denoted $\bm{x} \prec_w \bm{y}$, if~\cite{MarshallBook}
\begin{equation}
    \sum_{i=1}^k x_i \leq \sum_{i=1}^k y_i \quad \forall \; k=1,2,\dots,n.
\end{equation}
If in addition the total sum is the same,
\begin{equation}
    \sum_{i=1}^n x_i = \sum_{i=1}^n y_i,
\end{equation}
then we say that $\bm{y}$ \textbf{majorises} $\bm{x}$, denoted $\bm{x} \prec \bm {y}$.

The transformation laws for $M$ and $A$ now differ from each other and from the non-catalytic case.
A weak catalyst on its own, without invoking a thermal bath, is also able to effect change on $\bm{\mu}$:
\begin{theorem} \label{thm:M_weak_cat}
    \begin{enumerate}
        \item A weak catalytic transformation from $M$ to $M'$ is possible without use of a thermal bath if and only if
        \begin{equation}
            \bm{\mu'} \prec \bm{\mu}.
        \end{equation}
        If the system has $n$ modes, then the catalyst only needs at most $n-1$ modes.
        \item Allowing for the use of a thermal bath, the transformation is possible if and only if
        \begin{equation}
            \bm{{\mu'}^+} \prec_w \bm{\mu^+} \text{ and } \bm{{\mu'}^-} \prec_w \bm{\mu^-}.
        \end{equation}
        The catalyst only needs at most $n-1$ modes, and the bath at most $n$ modes.
    \end{enumerate}
\end{theorem}
See the proof in Appendix~\ref{app:M_weak_cat}.
Note that no statement can be made about the \emph{number} of positive or negative principal mode temperatures.
It is also worth noting that the construction of the operation in part (2) shows that we may interact the system with just the catalyst first, and then just with the bath as in the single-mode case.\\

For the parameters $\bg{\alpha}$, we find instead that the same set of transformations is possible, whether or not a thermal bath is invoked:
\begin{theorem} \label{thm:A_weak_cat}
    A weak catalytic transformation from $A$ to $A'$ is possible either with or without use of a thermal bath if and only if
    \begin{equation}
        \bm{\alpha'} \prec_w \bm{\alpha}.
    \end{equation}
    If the system has $n$ modes, we require at most $2n-1$ catalyst modes, or otherwise $n$ catalyst modes plus $n-1$ bath modes.
\end{theorem}
See the proof in Appendix~\ref{app:A_weak_cat}.
The key additional transformation provided by a weak catalyst is an operation on a given pair of modes.
For instance, the result on the $M$ matrix is a so-called T-transform~\cite[Lemma 2.B.1]{MarshallBook} on its eigenvalues, resulting in $\mu'_1 = (1-t) \mu_1 + t \mu_2$, $\mu'_2 = (1-t)\mu_2 + t \mu_1$ for some $t \in [0,1]$.

Again, it is important to note that the above conditions are only \emph{necessary conditions} for state transformation due to the fact that $M$ and $A$ are in general not simultaneously diagonalisable. Also, the derived multi-mode conditions do not fully recover the single-mode results presented in Section~\ref{sec:single_mode}\;--\;we are missing the conditions on the ratio between $\mu$ and $\alpha$. It is an open question to see whether we can derive a generalised version of the conditions on the ratio in the multi-mode case. 

\section{Physical implications}\label{sec:physical_implications}

\begin{table}[h]
    \bgroup
    \def\arraystretch{1.5}
    \setlength\tabcolsep{5pt}
    \begin{tabular}{c | c c}
        Transformation type        & \multicolumn{2}{c}{Ordering} \\ \hline
        No catalyst                & $\bm{\mu'}^\pm \leq \bm{\mu}^\pm,$    & $\bm{\alpha'} \leq \bm{\alpha}$  \\
        Strong catalyst            & $\bm{\mu'}^\pm \leq \bm{\mu}^\pm,$    & $\bm{\alpha'} \leq \bm{\alpha}$  \\
        Weak catalyst, no bath     & $\bm{\mu'} \prec \bm{\mu},$           & $\bm{\alpha'} \prec_w \bm{\alpha}$ \\
        Weak catalyst, with bath   & $\bm{\mu'}^\pm \prec_w \bm{\mu}^\pm,$ & $\bm{\alpha'} \prec_w \bm{\alpha}$ 
    \end{tabular}
    \egroup
    \caption{Summary of ordering relations on the $\bm{\mu}$ and $\bm{\alpha}$ parameters for different types of GTO transformations.}
    \label{tab:summary}
\end{table}

\begin{figure}[h]
    \includegraphics[width=0.4\textwidth]{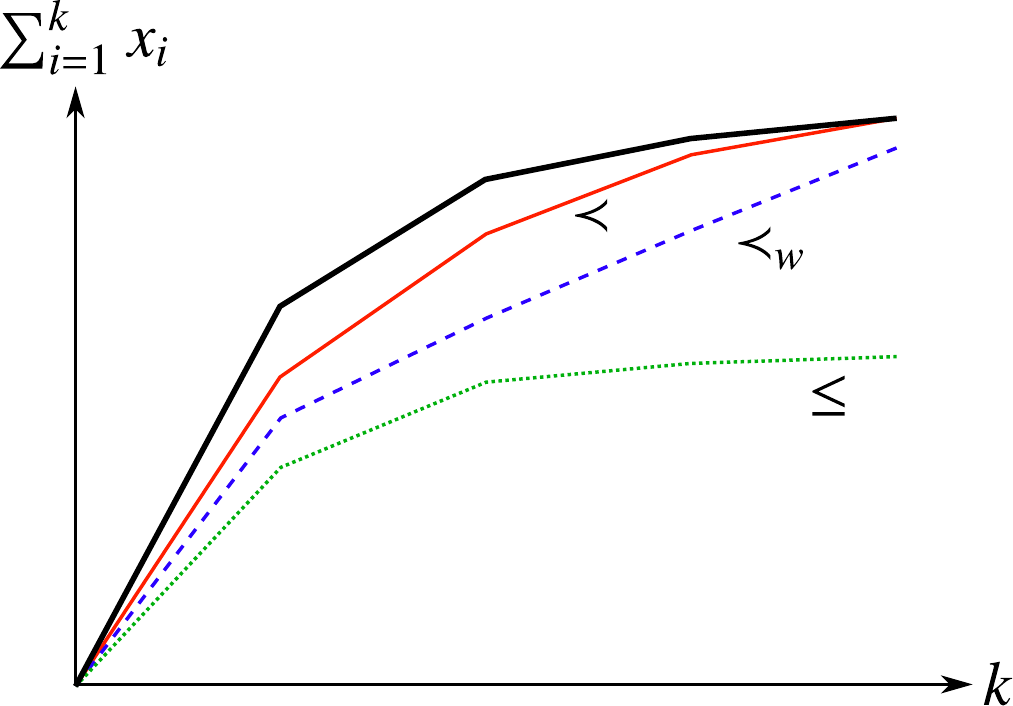}
    \caption{Lorenz curves (i.e., partial sums) demonstrating the different kinds of orderings encountered here. If the top curve (solid, black) is $\bm{x}$, then the lower curves from top to bottom satisfy respectively $\bm{x'} \prec \bm{x}$ (solid, red), $\bm{x'} \prec_w \bm{x}$ (dashed, blue), and $\bm{x'} \leq \bm{x}$ (dotted, green).}
    \label{fig:orders}
\end{figure}

\subsection{Concentration of athermal resources}
Here we discuss the physical meaning of the monotonicity results found for the parameters $\bm{\mu}$ and $\bm{\alpha}$; a summary of the different cases is given in Table~\ref{tab:summary}.

First consider the matrix $M$: given a choice of orthogonal modes, the diagonals $M_{ii}$ in the corresponding basis are related to the mean energy in each mode (recall that we assume vanishing first moments).
The principal mode temperatures $\mu_i$ are the diagonals in a basis where $M$ is diagonal.
In particular, these bases coincide if (but not only if, due to the other matrix $A$) the chosen modes are uncorrelated.
Due to the majorisation relation between the eigenvalues of a matrix and its diagonals, the distribution of $M_{ii}$ is generally more uniform than that of $\mu_i$.
For a set of initially uncorrelated modes isolated from the thermal environment and which interact with a passive linear unitary, this relation describes an approach to equilibrium via what could be described as heat exchange, since these components of energy are related to the quadrature fluctuations (rather than first moments).
The $\mu_i$ are intrinsic to the closed system, so are unchanged over this unitary evolution.

An important common feature of the relations in Table~\ref{tab:summary} is a restriction on the concentration of resources.
For instance, all the orderings $\leq, \prec, \prec_w$ require $\mu'^+_1 \leq \mu^+_1$.
This means it is impossible to concentrate the energy of multiple modes in such a way that the ``hottest" mode becomes hotter.
The same rule applies inversely to modes below the background temperature, so the coldest sub-thermal mode cannot be cooled.
Therefore an absorption refrigerator is an impossible machine within the Gaussian framework. This observation was also made in Refs.~\cite{Martinez2013Dynamics,Freitas2017Fundamental,Opatrny2021Nonlinear} -- however, we also see that further, more subtle conditions apply to concentration into larger subsets of modes.
For instance, either with no catalyst or with a strong catalyst, $\mu'^+_2 \leq \mu^+_2$ so the second hottest mode also cannot become hotter.
A weak catalyst instead opens up the possibility of heating this mode, subject to the constraint that the two hottest modes in total do not heat up, $\mu'^+_1 + \mu'^+_2 \leq \mu^+_1 + \mu^+_2$ -- that is, energy in the first mode can be traded for heating of the second mode.

Note that the case of a weak catalyst without interaction with a thermal bath makes no distinction between super- and sub-thermal modes; this is to be expected since there is no background reference temperature in this case.
The effect of including a thermal bath is then to separate the conditions on super- and sub-thermal modes, and also to allow for losses, in the sense that the total $\sum_{i=1}^{n_+} \mu^+_i$ can decrease, for instance.
So the super-thermal modes can in total lose energy to the bath, and the sub-thermal modes can in total gain energy from the bath.

Similar considerations apply to the principal mode asymmetries $\alpha_i$.
A notable difference compared with $\mu_i$ is that the ``lossy" property exists for a weak catalyst with no bath interaction.
Intuitively, this is due to the existence of operations that result in components of the asymmetry being transferred to correlations between the system and the catalyst.

\subsection{General limitations of Gaussian systems}
In addition to the constraints described above, there is a deeper limitation preventing Gaussian systems from making useful thermodynamical machines.
In order to develop this argument, we first need to understand how to describe work in the GTO framework.
In the discrete thermal operations resource theory~\cite{Horodecki2013Fundamental,Brandao2015Second}, a work battery is typically defined as a system that transitions from one pure energy eigenstate to another so that it does not change entropy, and a definite, non-fluctuating amount of work is exchanged with other systems\footnote{Generalisations do exist, for example see Ref.~\cite{HindsMingo2019Decomposable}}.

This definition cannot be used in GTOs because no energy eigenstate other than the vacuum is Gaussian.
Given the significance of quadratures in the Gaussian setting, it seems reasonable to instead define a \textbf{Gaussian work battery} as a system that transitions from one Gaussian state to another under the action of a displacement, $D(\bm{\xi}) = e^{\bm{\xi}\cdot \bm{a}^\dagger - \bar{\bm{\xi}} \cdot \bm{a}}$ -- so only its first moments can change, but not its CM.
The work is then defined as the change in mean energy of the battery.
Apart from the change in first moments, the energy distribution changes too, so this is necessarily a weakening of the usual requirements.
However, its entropy is a function of the CM and so is constant.

Depending on whether the work battery is allowed to become correlated with the system, it therefore functions at the level of the CM as either a strong or a weak catalyst.
Either way, the final CM of the system is only a function of the initial CM and completely independent of the first moments.
Thus, the possible transformations of the system's CM are the same no matter how much energy is stored in the battery -- work has no effect.
In this Gaussian setting, therefore, no useful machine can be constructed whereby work is traded for resources at the CM level.

What, then, can a Gaussian work battery do?
Only non-trivial transformations involving first moments alone are possible.
In general, given a system $\mrm{S}$ and a work battery $\mrm{W}$ with first moments $\bm{r}_\mrm{S}$ and $\bm{r}_\mrm{W}$, a joint passive linear unitary represented by $S \in \mrm{K}(n_\mrm{S} + n_\mrm{W})$ has the action
\begin{equation}
    \begin{pmatrix}
        \bm{r'}_\mrm{S} \\ \bm{r'}_\mrm{W}
    \end{pmatrix} = S
    \begin{pmatrix}
        \bm{r}_\mrm{S} \\ \bm{r}_\mrm{W}
    \end{pmatrix},
\end{equation}
i.e., the point in the combined phase space is simply rotated about the origin by an orthogonal symplectic matrix.
For example, with one mode each, a beam-splitter with reflectivity $r$ has the effect on a pair of coherent states $\ket{\eta}_\mrm{S} \ket{\zeta}_\mrm{W} \to \ket{r \eta + \sqrt{1-r^2} \zeta}_\mrm{S} \ket{r \zeta - \sqrt{1-r^2}\eta}_\mrm{W}$.
One could view this as a quasi-deterministic exchange of energy.
In general, a displacement can be applied to $\mrm{S}$ by choosing a battery $W$ with the same CM as $\mrm{S}$ and suitably chosen first moments along with a set of beam-splitters.
So, while displacements are not free within GTO, they can be accomplished with an additional system that exchanges energy in the form of first moments but is strongly catalytic in its second moments.
Alternatively, if we do not want to adapt $\mrm{W}$ to fit the CM of $\mrm{S}$, then we can use an arbitrary CM and a large displacement $\zeta$ -- in the limit $\abs{\zeta} \to \infty$ with $r \abs{\zeta}$ held constant, a displacement is performed on $\mrm{S}$~\cite{Paris1996Displacement}. For finite $\zeta$, this is an instance of an approximate strong catalytic transformation, so Theorem~\ref{thm:multi_strong_cat} guarantees that the thermal monotone laws found here still hold approximately.

At a broader level, it is also worthwhile noting that the direct-sum structure of multi-mode Gaussian systems seems crucial for these general limitations in the Gaussian setting. The main reason why introducing catalysts does not have as dramatic an impact on the ordering relations of $\bm{\mu}$ and $\bm{\alpha}$ as in other resource theories is that catalysts do not change the pre-existing elements of $\bm{\mu}$ and $\bm{\alpha}$ but just add more elements to such vectors. In contrast, in the tensor-product structure of the full Hilbert space, the catalyst's vector elements are multiplied with the system's elements, which can significantly affect vector ordering. The direct-sum structure of multi-mode Gaussian systems plays an important role in other limitations to Gaussian resource theories as well; for example, in Ref.~\cite{Lami2018Gaussian} it is shown that the tensorisation property of a resource monotone, due to the direct-sum structure of CMs, results in a no-go theorem for Gaussian resource distillation.

%%%%%%%%%%%%%%
\section{Conclusions}\label{sec:conclusions}
In this paper, we look at possible state transformations under Gaussian thermal operations when catalysts are allowed. We define two different types of catalysts: strong catalysts which must not only come back to the original state but also end up uncorrelated to the system, and weak catalysts whose local state only needs to be reset. We ask whether exploiting these catalysts permits more state transformations than the non-catalytic case. Despite the general benefit of catalysts in other contexts, we find that strong catalysts do not enable more state transformations, while weak catalysts can achieve more but limited state transformations compared to the non-catalytic case. Alongside with the principal mode temperatures introduced in \cite{Narasimhachar2021Thermodynamic}, we define another resource monotone, the principal mode asymmetry, to describe the state transition conditions. We fully characterise the necessary and sufficient conditions for state transformation under single-mode catalytic GTOs for both types of catalysts, and also provide new necessary conditions for the multi-mode case in terms of principal mode temperatures and symmetries. We discuss physical implications of our results on Gaussian thermal machines such as Gaussian refrigerators and Gaussian work batteries.

In general, due to the stringent requirement imposed by the direct sum composition of phase spaces, resource theories in the Gaussian regime allow for very limited operations. It is therefore natural to inject non Gaussian elements into the theory. 
In this regard, it would be interesting to see what transformations would be unlocked if Gaussian thermal operations are combined with resourceful non-Gaussian catalysts. The conditions determined in this study would still hold at the level of second moments (since Gaussian states allow one to reproduce all physical CMs and their transformations would still be as described here), but more interesting dynamics could be allowed at the full Hilbert space level. 
More fundamentally, however, non-Gaussian operations need to be included as resources to break the constraints we have found on the manipulation of second moments.

\section*{Acknowledgments}
This project has received funding from the European Union's Framework Programme for Research and Innovation Horizon 2020 (2014-2020) under the Marie Sk\l odowska-Curie Grant Agreement No.\ 945422.
B.\ Y.\ was also supported by grant number (FQXi FFF Grant number FQXi-RFP-1812) from the Foundational Questions Institute and Fetzer Franklin Fund, a donor advised fund of Silicon Valley Community Foundation. H.\ H.\ J.\ is supported through a studentship in the Centre for Doctoral Training on Controlled Quantum Dynamics at Imperial College London funded by the EPSRC (EP/L016524/1).  G.\ A.\ acknowledges
financial support from FAPESP (Grant No.~2017/07973-5).

\appendix

\section{Proof of Theorem~\ref{thm:no_catalyst}} \label{app:no_catalyst}
    The initial covariance matrix is $\sigma_\mrm{S} \oplus \sigma_\mrm{B}$ where $\sigma_\mrm{B}$ is thermal, so $M_\mrm{SB} = M_\mrm{S} \oplus 0_\mrm{B}$.
    Partitioning the interacting passive linear unitary $U$ in this way,
    \begin{equation}
        U = \begin{pmatrix}
            P & Q \\ R & S
        \end{pmatrix}.
    \end{equation}
    The submatrix $P$ is a contraction, meaning that $P^\dagger P \leq \id, \, P P^\dagger \leq \id$.
    Partitioning the final matrix $M' = U M U^\dagger$ in the same way (with stars denoting unspecified elements),
    \begin{equation}
        M' = \begin{pmatrix}
            M'_\mrm{S} & * \\ * & *
        \end{pmatrix},
    \end{equation}
    we find that $M'_\mrm{S} = P M_\mrm{S} P^\dagger$.
    An extension of Ostrowski's theorem~\cite[Corollary 4.5.11]{Horn1985Matrix} says that there exist $r_i \in [p_n^2, p_1^2]$, where $p_1 \geq \dots \geq p_n$ are the singular values of $P$, such that $\mu'_i = r_i \mu_i$.
    The necessity of the conditions in part (1) follows from $P$ being a contraction, so that $r_i \in [0,1]$.

    Sufficiency is seen from the fact that $\bm{\mu'}$ can be obtained from $\bm{\mu}$ by multiplying each element by some $r \in [0,1]$.
    We call such an operation on each element an \emph{L-transform}.
    This can be performed on each mode by interacting with a single thermal mode via a beam splitter: concentrating on just this pair of modes, take
    \begin{equation}
        U = \begin{pmatrix}
            \sqrt{r} & -\sqrt{1-r} \\ \sqrt{1-r} & \sqrt{r}
        \end{pmatrix}.
    \end{equation}
    Without loss of generality, we can assume the initial $M$-matrix of the system to be diagonal, so for this pair of modes,
    \begin{equation}
        M = \begin{pmatrix}
            \mu & 0 \\ 0 & 0
        \end{pmatrix} \longrightarrow M' = UMU^\dagger =
        \begin{pmatrix}
            r \mu & -\sqrt{r(1-r)} \mu \\ -\sqrt{r(1-r)} \mu & (1-r)\mu
        \end{pmatrix},
    \end{equation}
    and tracing out the bath just gives the diagonal element $r \mu$.\\

    For part (2), we similarly have $A' = U (A_\mrm{S} \oplus 0_\mrm{B}) U^T$ so that $A'_\mrm{S} = P A_\mrm{S} P^T$.
    A related result for complex symmetric matrices~\cite[Theorem 4.5.13]{Horn1985Matrix} says that $\alpha'_i = r_i \alpha$, where $r_i \in [q_n, q_1]$ and $q_1 \geq \dots \geq q_n$ are the eigenvalues of $PP^\dagger$ -- which again lie in the interval $[0,1]$.
    Sufficiency follows from starting in a basis with $A = \mrm{diag}(\bm{\alpha})$ and then applying the same $U$ as above for each mode.
    The calculation proceeds identically since all matrix elements are real.

\section{Proof of Theorem~\ref{thm:single_mode_weak_cat_GTO}}\label{app:single_mode_weak_cat_GTO}
    In this section, we will prove Theorem~\ref{thm:single_mode_weak_cat_GTO}, the state transition conditions under single-mode weak catalytic GTOs. We prove the conditions with weak catalysts first as the proof for the conditions with strong catalysts borrows a lot of results from this case.
\begin{proof}    
    \textit{(Necessary conditions)} 
    We start from Eq.~\eqref{eq:MandA_transformation_via_P},
    \begin{align}
    M_\mrm{SC}'=\begin{pmatrix}
    |p_{11}|^2\mu_\mrm{S}+|p_{12}|^2\mu_\mrm{C} & p_{11}p_{21}^*\mu_\mrm{S}+p_{12}p_{22}^*\mu_\mrm{C} \\
    p_{11}^*p_{21}\mu_\mrm{S}+p_{12}^*p_{22}\mu_\mrm{C} & |p_{21}|^2\mu_\mrm{S}+|p_{22}|^2\mu_\mrm{C}
    \end{pmatrix} , \;
    A_\mrm{SC}'=\begin{pmatrix}
    p_{11}^2\alpha_\mrm{S}+p_{12}^2\alpha_\mrm{C} & p_{11}p_{21}\alpha_\mrm{S}+p_{12}p_{22}\alpha_\mrm{C} \\
    p_{11}p_{21}\alpha_\mrm{S}+p_{12}p_{22}\alpha_\mrm{C} & p_{21}^2\alpha_\mrm{S}+p_{22}^2\alpha_\mrm{C}
    \end{pmatrix},
    \end{align}
    where $p_{ij}=(P)_{ij}$ are the elements of the sub-matrix $P$. The final state is described by
    \begin{subequations}\label{eq:single_mode_final}
    \begin{align}
        \label{eq:single_mode_final_mu}
        \mu_\mrm{S}'& =|p_{11}|^2\mu_\mrm{S}+|p_{12}|^2\mu_\mrm{C}\, \\
        \label{eq:single_mode_final_alpha}
        \alpha_\mrm{S}' & = p_{11}^2\alpha_\mrm{S}+p_{12}^2\alpha_\mrm{C}\,.
    \end{align}
    \end{subequations}
    We want to find the relations between the initial parameters $\mu_\mrm{S}$, $\alpha_\mrm{S}$ and the final parameters $\mu_\mrm{S}'$ and $\alpha_\mrm{S}'$. To do so, we need to exploit the \emph{catalytic conditions} given by
    \begin{subequations}
    \label{eq:catalytic_conditions}
    \begin{align}
        |p_{21}|^2\mu_\mrm{S}+|p_{22}|^2\mu_\mrm{C} & = \mu_\mrm{C}\,, \\
        p_{21}^2\alpha_\mrm{S}+p_{22}^2\alpha_\mrm{C} & = \alpha_\mrm{C}\,.
    \end{align}
    \end{subequations}
    Firstly, let us assume $|p_{22}|^2=1$. This is a trivial case as it implies $p_{12}=p_{21}=0$, so that $\mu_\mrm{S}'=|p_{11}|^2\mu_\mrm{S}$ and $\alpha_\mrm{S}'=p_{11}^2\alpha_\mrm{S}$ which satisfies the conditions in Eq.~\eqref{eq:single_mode_weak_catGTO}. In the following, we will assume $|p_{22}|^2\neq1$.
    
    Assuming $\mu_S\neq0$ and $\alpha_S\neq0$, we can rewrite the catalytic conditions in Eq.~\eqref{eq:catalytic_conditions} as
    \begin{align}
        \frac{\mu_\mrm{C}}{\mu_\mrm{S}} = \frac{|p_{21}|^2}{1-|p_{22}|^2}\,, \quad \frac{\alpha_\mrm{C}}{\alpha_\mrm{S}}=\frac{p_{21}^2}{1-p_{22}^2}\,,
    \end{align}
    which implies that $\frac{p_{21}^2}{1-p_{22}^2}$ is real and non-negative\footnote{Recall the parameters $\alpha$ are real and non-negative -- see Section~\ref{subsec:PMTemperatures_PMAsymmetries}.}.
    Substituting these to the expressions of the final state in Eq.~\eqref{eq:single_mode_final} obtains
    \begin{subequations}
    \begin{align}
    \mu_\mrm{S}' &= |p_{11}|^2 \, \mu_\mrm{S} + |p_{12}|^2 \, \mu_\mrm{C} = \left(|p_{11}|^2 + \frac{|p_{12}|^2|p_{21}|^2}{1-|p_{22}|^2}\right)\mu_\mrm{S} \,, \label{eq:muSprime_in_terms_of_muS}\\
    \alpha'_\mrm{S} &= p_{11}^2 \, \alpha_\mrm{S} + p_{12}^2 \, \alpha_\mrm{C} = \left( p_{11}^2 + \frac{p_{12}^2p_{21}^2}{1-p_{22}^2}\right) \alpha_\mrm{S} \,,
    \end{align}
    \end{subequations}
    which again implies that both expressions in the brackets are real.
    Then, we have
    \begin{align*}
        p_{11}^2 + \frac{p_{12}^2 p_{21}^2}{1-p_{22}^2} &\leq \abs{p_{11}^2 + \frac{p_{12}^2 p_{21}^2}{1-p_{22}^2}}  \leq \abs{p_{11}^2} + \abs{\frac{p_{12}^2p_{21}^2}{1-p_{22}^2}} \\
        &= \abs{p_{11}}^2 + \abs{p_{12}}^2 \frac{\abs{p_{21}^2}}{\abs{1-p_{22}^2}}
        \leq \abs{p_{11}}^2 + \frac{\abs{p_{21}}^2\abs{p_{12}}^2}{1-\abs{p_{22}}^2}\,,
    \end{align*}
    where we used the reverse triangle inequality, $\abs{x-y}\geq\abs{\abs{x}-\abs{y}}$, and $\abs{p_{22}}\leq1$ in the last inequality. This shows $\alpha_\mrm{S}'/\alpha_\mrm{S}\leq\mu_\mrm{S}'/\mu_\mrm{S}$. 

    The remaining part is to prove $\mu_\mrm{S}'/\mu_\mrm{S}\leq1$. This can be easily shown using the condition of the $P$ matrix $\id-PP^{\dagger}\geq 0$. We have that
    \begin{equation}
        \id - PP^{\dagger} = 
        \begin{pmatrix}
        1-|p_{11}|^2-|p_{12}|^2 & -p_{11}p_{21}^*-p_{12}p_{22}^* \\
        -p_{11}^*p_{21}-p_{12}p_{22}^* & 1-|p_{21}|^2-|p_{22}|^2
        \end{pmatrix}
        \geq0,
    \end{equation}
    which implies 
    \begin{subequations}\label{eq:PP_ddager_condition}
    \begin{align}
        \label{eq:PPdagger_1} 1-|p_{11}|^2-|p_{12}|^2\geq0\,,\\
        \label{eq:PPdagger_2} 1-|p_{21}|^2-|p_{22}|^2\geq0\,.
    \end{align}
    \end{subequations}
    Then, we obtain
    \begin{align*}
        |p_{11}|^2 + \frac{|p_{12}|^2|p_{21}|^2}{1-|p_{22}|^2} \leq |p_{11}|^2 + \frac{(1-|p_{11}|^2)(1-|p_{22}|^2)}{1-|p_{22}|^2} = 1\,,
    \end{align*}
    where we used Eq.~\eqref{eq:PP_ddager_condition} in the first inequality. This completes the proof.
    
    When $\mu_\mrm{S}=0$, the catalytic condition implies that either $|p_{22}|^2=1$ or $\mu_\mrm{C}=0$. If $|p_{22}|^2=1$, from the earlier argument on the case when $|p_{22}|^2=1$, we have $\mu_\mrm{S}'=0$ and $\alpha_\mrm{S}'=p_{11}^2\alpha_\mrm{S}\leq\alpha_\mrm{S}$. If $\mu_\mrm{C}=0$, then $\mu_\mrm{S}'=0$, and all the results on $\alpha_\mrm{S}'$ are same as above; $\alpha_\mrm{S}'\leq\alpha_\mrm{S}$. A similar argument works for the case when $\alpha_\mrm{S}=0$.

    \textit{(Sufficient conditions)} We would like to show that for any $\mu_\mrm{S}'$ and $\alpha_\mrm{S}'$ satisfying Eq.~\eqref{eq:single_mode_weak_catGTO}, we can find a weak catalytic GTO transformation which maps the initial state to the final state described by $\mu_\mrm{S}'$ and $\alpha_\mrm{S}'$. To do so, we provide an explicit weak catalytic GTO transformation which can map the initial state with $\mu_\mrm{S}$ and $\alpha_\mrm{S}$ to the final state with $\mu_\mrm{S}'$ and $\alpha_\mrm{S}'$.
    Consider a weak catalytic GTO transformation consisting of the following two steps:
    \begin{enumerate}
        \item By coupling the system with the catalyst, reduce the $\alpha_\mrm{S}$ to $\bar{\alpha}_\mrm{S}=(\mu_\mrm{S}/\mu_\mrm{S}')\alpha_\mrm{S}'$ whilst keeping the same $\mu_\mrm{S}$. This can be done with the catalyst described by $\mu_\mrm{C} = \mu_\mrm{S}$ and $\alpha_\mrm{C} = \frac{\alpha_\mrm{S}-\bar{\alpha}_\mrm{S}}{2}$, and the following unitary operation $U_1$ acting on the matrices $M$ and $A$:
        \begin{align} \label{eqn:alpha_unitary}
            U_1=\begin{pmatrix}
            \sqrt{a} & i\sqrt{1-a} \\
            \sqrt{1-a} & -i\sqrt{a}
            \end{pmatrix}\,,
        \end{align}
        where $a=\frac{\alpha_\mrm{S}+\bar{\alpha}_\mrm{S}}{3\alpha_\mrm{S}-\bar{\alpha}_\mrm{S}}\in[\frac{1}{3},1]$. $U_1$ can be realised by passive unitaries on the system and the catalyst; coupling the system and the catalyst via a beam splitter with transmissivity $a$ together with a phase shift.
        \item Couple the system with the thermal bath and thermalise the system to the final state with $\mu_\mrm{S}'$ and $\alpha_\mrm{S}'$. This can be done by applying a beam splitter with transmissivity  $\mu_\mrm{S}'/\mu_\mrm{S}$ to the bath and the system, which is now described by $\mu_\mrm{S}$ and $\bar{\alpha}_\mrm{S}$, and tracing out the bath mode.
    \end{enumerate}
    It is not difficult to check that above two steps transform $\mu_\mrm{S}, \alpha_\mrm{S}$ to $\mu_\mrm{S}', \alpha_\mrm{S}'$, and the whole process is a GTO transformation. Geometrically, the first step describes moving vertically (reducing the asymmetry $\alpha$) in Fig.~\ref{fig:single_mode_mu_alpha}, and the second step is moving towards to the origin (pure thermalisation)\footnote{This is because a thermal state at the background temperature is always described by $M=A=0$. See Section~\ref{subsec:PMTemperatures_PMAsymmetries}.}.
\end{proof}

\section{Proof of Theorem~\ref{thm:single_mode_strong_cat_GTO}} \label{app:single_mode_strong_cat_GTO}
    In this appendix, we will derive the necessary and sufficient conditions for state transition under single-mode strong catalytic GTOs. We will borrow a few results from the proof of Theorem~\ref{thm:single_mode_weak_cat_GTO} for the case of weak catalysts, which is derived in Appendix~\ref{app:single_mode_weak_cat_GTO}.
\begin{proof}
    \textit{(Necessary conditions)} We start again from Eq.~\eqref{eq:MandA_transformation_via_P}, and the final state is described by
    \begin{subequations}
    \label{eq:final_stat_in_terms_of_P}
    \begin{align}
        \mu_\mrm{S}' & =|p_{11}|^2\mu_\mrm{S}+|p_{12}|^2\mu_\mrm{C}\, \\
        \alpha_\mrm{S}' & = p_{11}^2\alpha_\mrm{S}+p_{12}^2\alpha_\mrm{C}\,.
    \end{align}
    \end{subequations}
    Again, we want to find the relations between the initial parameters $\mu_\mrm{S}, \alpha_\mrm{S}$ and the final parameters $\mu_\mrm{S}', \alpha_\mrm{S}'$. This time, we can exploit not only the catalytic conditions in Eq.~\eqref{eq:catalytic_conditions} but also the \emph{no-correlation conditions} given by vanishing off-diagonal elements,
    \begin{subequations}\label{eq:no_correlation_strongCat}
    \begin{align}
        p_{11}p_{21}^*\mu_\mrm{S}+p_{12}p_{22}^*\mu_\mrm{C} = 0\,, \\
        p_{11}p_{21}\alpha_\mrm{S}+p_{12}p_{22}\alpha_\mrm{C} = 0\,.
    \end{align}
    \end{subequations}
    As we are looking at a more restricted case (with more conditions) than the one with weak catalysts, the conditions for weak catalysts must hold regardless of the no-correlation conditions\;--\;the condition in Eq.~\eqref{eq:single_mode_weak_catGTO} automatically holds due to the proof of Theorem~\ref{thm:single_mode_weak_cat_GTO} in the last appendix. Now, we just need to show that the ratios $p=\frac{\mu_\mrm{S}'}{\mu_\mrm{S}}$ and $q=\frac{\alpha_\mrm{S}'}{\alpha_\mrm{S}}$ are the same. 
    
    Firstly, let us assume $\mu_\mrm{S}\neq0$ and $\alpha_\mrm{S}\neq0$. We notice from the no-correlation conditions in Eq.~\eqref{eq:no_correlation_strongCat} that
    \begin{align}\label{eq:singleMode_strongCat_no_corr_conditions_1}
        \frac{\mu_\mrm{C}}{\mu_\mrm{S}}=-\frac{p_{11}p_{21}^*}{p_{12}p_{22}^*} \in \mathbb{R}\,, \quad \frac{\alpha_\mrm{C}}{\alpha_\mrm{S}} = -\frac{p_{11}p_{21}}{p_{12}p_{22}} \in \mathbb{R}\,,
    \end{align}
    which implies that $\mu_\mrm{C}/\mu_\mrm{S}=\pm \alpha_\mrm{C}/\alpha_\mrm{S}$.
    Combining this result with the catalytic conditions in Eq.~\eqref{eq:catalytic_conditions} gives us
    \begin{align}
        \frac{\mu_\mrm{C}}{\mu_\mrm{S}} = \frac{|p_{21}|^2}{1-|p_{22}|^2} = -\frac{p_{11}p_{21}^*}{p_{12}p_{22}^*} &\;\implies\; p_{11}=\frac{p_{21}p_{12}p_{22}^*}{|p_{22}|^2-1}\,, \label{eq:singleMode_catalytic_conditions_modified_mu}\\
        \frac{\alpha_\mrm{C}}{\alpha_\mrm{S}} = \frac{p_{21}^2}{1-p_{22}^2} = -\frac{p_{11}p_{21}}{p_{12}p_{22}} &\;\implies\; p_{11} = \frac{p_{21}p_{12}p_{22}}{p_{22}^2-1}\,, \label{eq:singleMode_catalytic_conditions_modified_alpha}
    \end{align}
    where we assume that $|p_{22}|\neq1$, $p_{12}\neq0$, and $p_{22}\neq0$. Then, using this together with the catalytic conditions, we can obtain from Eq.~\eqref{eq:single_mode_final} that
    \begin{align} \label{eqn:mu_ratio}
        \mu_\mrm{S}' &= \left(\frac{|p_{21}||p_{12}||p_{22}|}{|p_{22}|^2-1}\right)^2\mu_\mrm{S} + |p_{12}|^2\left(\frac{|p_{21}|^2}{1-|p_{22}|^2}\mu_\mrm{S}\right) \\
        &= \frac{|p_{21}|^2|p_{12}|^2|p_{22}|^2 + (1-|p_{22}|^2)|p_{12}|^2|p_{21}|^2}{(1-|p_{22}|^2)^2}\mu_\mrm{S}
        =\frac{|p_{12}|^2|p_{21}|^2}{(1-|p_{22}|^2)^2}\mu_\mrm{S}\,.
    \end{align}
    Also, we can obtain that
    \begin{align}
        \alpha_\mrm{S}' &= \left(\frac{p_{21}p_{12}p_{22}}{p_{22}^2-1}\right)^2\alpha_\mrm{S} + p_{12}^2\left(\frac{p_{21}^2}{1-p_{22}^2}\alpha_\mrm{S}\right)
        = \frac{p_{12}^2p_{21}^2}{(1-p_{22}^2)^2}\alpha_\mrm{S} = \frac{p_{12}^2}{p_{21}^2} \times\frac{p_{21}^4}{(1-p_{22}^2)^2}\alpha_\mrm{S}\,,
    \end{align}
    where we assume $p_{21}\neq0$ here\footnote{If $p_{21}=0$, then $|p_{22}|^2=1$ because of the catalytic conditions in Eq.~\eqref{eq:catalytic_conditions}, but we already assumed $|p_{22}|^2\neq1$.}.
    Since $p_{21}^2/(1-p_{22}^2)= \alpha_\mrm{C}/\alpha_\mrm{S}=\pm \mu_\mrm{C}/\mu_\mrm{S}=\pm  |p_{21}|^2/(1-|p_{22}|^2)\in\mathbb{R}$, this leads to
    \begin{align}
        \alpha_\mrm{S}'= \pm \frac{p_{12}^2}{p_{21}^2} \frac{|p_{21}|^4}{(1-|p_{22}|^2)^2}\alpha_\mrm{S} = \pm \abs{\frac{p_{12}^2}{p_{21}^2}}\frac{|p_{21}|^4}{(1-|p_{22}|^2)^2}\alpha_\mrm{S} = + \frac{|p_{12}|^2|p_{21}|^2}{(1-|p_{22}|^2)^2}\alpha_\mrm{S}\,,
    \end{align}
    where we used the fact that the factor must be real in the second equality, and we chose a $+$ sign instead of a $-$ sign at the end as $\alpha$ is always non-negative. This proves that $\alpha_\mrm{S}'/\alpha_\mrm{S}=\mu_\mrm{S}'/\mu_\mrm{S}$. 
    
    We have to look at the remaining cases: (i) When $p_{12}=0$, then either $p_{11}$ or $p_{21}$ must be zero because of the no-correlation conditions in Eq.~\eqref{eq:no_correlation_strongCat}. When $p_{12}=p_{11}=0$, $\mu_S'=\alpha_S'=0$, which means that the final state is thermal. When $p_{12}=p_{21}=0$, then $\mu_\mrm{S}'=|p_{11}|^2\mu_\mrm{S}$, and $\alpha_\mrm{S}'=|p_{11}|^2\alpha_\mrm{S}$ with $|p_{11}|^2\leq1$. (ii) When $p_{22}=0$, again either $p_{11}$ or $p_{21}$ must be zero. If $p_{22}=p_{11}=0$, $\mu_\mrm{S}'=|p_{12}|^2|p_{21}|^2\mu_\mrm{S}$, and $\alpha_\mrm{S}'=|p_{12}|^2|p_{21}|^2\alpha_\mrm{S}$ with $|p_{12}|^2|p_{21}|^2\leq1$. If $p_{22}=p_{21}=0$, the catalyst is thermal, resulting in a normal non-catalytic single-mode GTO transformation. (iii) When $|p_{22}|^2=1$, this implies $p_{12}=p_{21}=0$, so that $\mu_\mrm{S}'=|p_{11}|^2\mu_\mrm{S}$, and $\alpha_\mrm{S}'=|p_{11}|^2\alpha_\mrm{S}$ with $|p_{11}|^2\leq1$. All three cases satisfy the conditions in Theorem~\ref{thm:single_mode_strong_cat_GTO}.

    When $\mu_\mrm{S}=0$, from the catalytic conditions, we have either $|p_{22}|^2=1$ or $\mu_\mrm{C}=0$, and from the no-correlation conditions we have either $p_{12}=0$ or $p_{22}=0$ or $\mu_\mrm{C}=0$. It is not difficult to check that in all possible cases, it holds that $\mu_\mrm{S}'=0$ and $\alpha_\mrm{S}'\leq\alpha_\mrm{S}$, which satisfies the conditions. A similar argument also holds for the case when $\alpha_\mrm{S}=0$.
    
    \textit{(Sufficient conditions)}
    If we define $\gamma\leq1$ such that $|\mu_\mrm{S}'|=\gamma|\mu_\mrm{S}|$ and $|\alpha_\mrm{S}'|=\gamma|\alpha_\mrm{S}|$, the state transition condition can be expressed in terms of CMs as
    \begin{align}
        \sigma_\mrm{S}'=\gamma\sigma_\mrm{S}+(1-\gamma)\sigma_\mrm{B},
    \end{align}
    where $\sigma_\mrm{B}$ is the CM of the bath mode. Thus, the state transformation described by strong catalytic GTO transformations is mixing the initial CM of the system with the thermal bath, which can be achieved by a non-catalytic single-mode GTO transformation. As non-catalytic single-mode GTOs are included in single-mode strong catalytic GTOs, any final state satisfying the conditions in Eq.~\eqref{eq:single_mode_GTO_condition_in_mualpha} or Eq.~\eqref{eq:single_mode_GTO_condition_in_CM} can be also achieved by single-mode strong catalytic GTO transformations. 
\end{proof}

\section{Proof of Theorem~\ref{thm:multi_strong_cat}} \label{app:multi_strong_cat}
    For this proof, we make use of two intermediate results.

    \begin{lemma} \label{lem:approx_ordering}
        Let $\bm{z},\,\bm{z'}$ and $\bm{\tilde{z}}$ be non-increasingly ordered lists of length $l$ such that
        \begin{equation}
            \bm{\tilde{z}} \leq \bm{z} \quad \text{ and } \quad \sum_{i=1}^l \abs{\tilde{z}_i - z'_i} \leq \delta.
        \end{equation}
        Then
        \begin{equation}
            \sum_{i=1}^l \pos{z'_i - z_i} \leq \delta.
        \end{equation}
        \begin{proof}
            \begin{align}
                \sum_{i=1}^l \pos{z'_i - z_i} & = \sum_{i\colon z'_i > z_i} z'_i - z_i \nonumber \\
                    & \leq \sum_{i\colon z'_i > z_i} z'_i - \tilde{z}_i \nonumber \\
                    & \leq \sum_{i\colon z'_i > z_i} \abs{z'_i - \tilde{z}_i} \nonumber \\
                    & \leq \sum_{i=1}^l \abs{z'_i - \tilde{z}_i} \nonumber \\
                    & \leq \delta.
            \end{align}
        \end{proof}
    \end{lemma}
    
    \begin{theorem} \label{thm:approx_increases}
        Let $\bm{x} = (x_1,\dots,x_n)$, $\bm{x'} = (x'_1,\dots,x'_n)$ and $\bm{y} = (y_1,\dots,y_m)$ be non-increasing lists of real numbers, and define the composite ordered lists $\bm{z} = (\bm{x},\bm{y})^\downarrow$, $\bm{z'} = (\bm{x'},\bm{y})^\downarrow$.
        Suppose that there exists $\bm{\tilde{z}}$ such that
        \begin{equation}
            \bm{\tilde{z}} \leq \bm{z}  \quad \text{ and } \quad \sum_{i=1}^{n+m} \abs{\tilde{z}_i - z'_i} \leq \delta.
        \end{equation}
        Then
        \begin{equation}
            \sum_{i=1}^n \pos{x'_i - x_i} \leq \delta.
        \end{equation}
        \begin{proof}
            The main useful idea is to construct a partition of the indices of $\bm{z},\, \bm{z'}$ into contiguous blocks $B_1,B_2,\dots$ such that in each block, the $x_i$ values either all increase or all decrease when going from $\bm{z}$ to $\bm{z'}$ -- see Fig.~\ref{fig:blocks} for an illustration.
            This will let us separate out those values that break the monotonicity condition and bound how much they do so.
            Let $p(x_i)$ denote the index position of $x_i$ in $\bm{z}$, and similarly $p'(x'_i)$ for $x'_i$ in $\bm{z'}$.
            Note that a unique designation of indices is possible if we adopt the convention that, if $y_j = x_i$, then $y_j$ appears before $x_i$ in the list $\bm{z}$.

            \begin{figure*}[h]
                \includegraphics[width=0.35\textwidth]{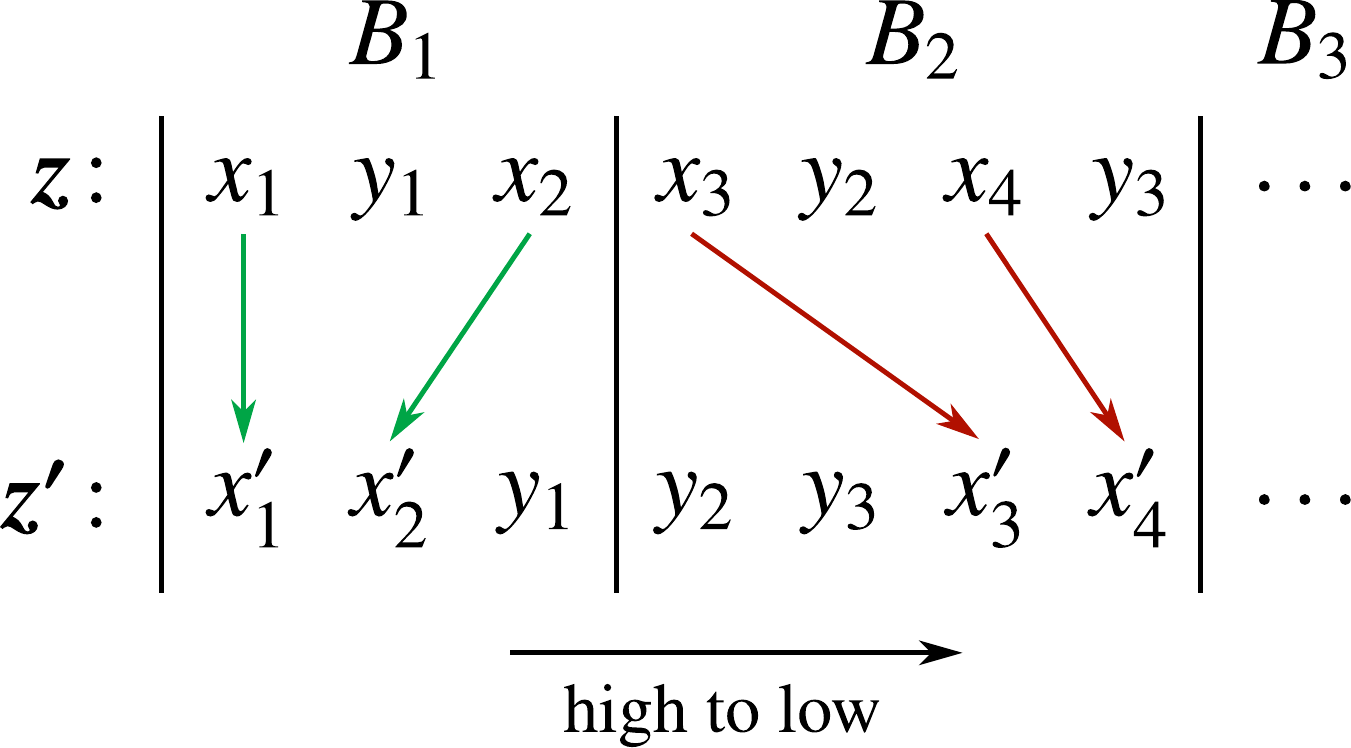}
                \caption{Illustration of the block partitioning of non-increasingly ordered lists $\bm{z} = (\bm{x},\bm{y})^\downarrow$, $\bm{z'}=(\bm{x'},\bm{y})^\downarrow$. In this case, $B_1$ is an increasing block, $B_2$ is decreasing, and so on alternately. Therefore $x'_1 > x_1$, $x'_2 > x_2$, $x'_3 \leq x_3$ and $x'_4 \leq x_4$.}
                \label{fig:blocks}
            \end{figure*}

            The end result of all this is that the set of indices $i$ of all $x_i$ in $B_b$ and $x'_i$ in $B_b$ are the same for each block $B_b$.
            Within an ``increasing" block, $x'_i > x_i$ and within a ``decreasing" block, $x'_i \leq x_i \; \forall i$.
            This claim will be proved separately below by showing how to construct the blocks.
            Using this, it is easy to see that the set of $y_j$ is also the same in each corresponding block.
            In other words, to send $\bm{z} \to \bm{z'}$, in each block we apply a permutation of the elements and replace $x_i \to x'_i$.
    
            Since the $y_j$ values match up in each block $B_b$, we have
            \begin{align}
                \sum_{k \in B_b} z'_k - z_k & = \sum_{i\colon p(x_i) \in B_b} x'_i - x_i. \\
            \end{align}
            Summing over all increasing blocks and using Lemma~\ref{lem:approx_ordering},
            \begin{align}
                \sum_{i=1}^n \pos{x'_i - x_i} & = \sum_{B_b \text{ increasing}} \; \sum_{i\colon p(x_i) \in B_b} x'_i - x_i \nonumber \\
                    & = \sum_{B_b \text{ increasing}} \; \sum_{k \in B_b} z'_k - z_k \nonumber \\
                    & \leq \sum_{k=1}^{n+m} \pos{z'_k - z_k} \nonumber \\
                    & \leq \delta\,.
            \end{align}

            \emph{Construction of the block partition:}\\
    
            Starting from the beginning of the list, suppose that $x'_1 > x_1$, and find the lowest $j$ such that $x'_j \leq x_j$.
            The list $\{1,2,\dots,p(x_{j-1})\}$ is then called the first block $B_1$ -- this is an ``increasing" block, such that $x'_i > x_i \; \forall p(x_i) \in B_1$.
            If instead $x'_1 \leq x_1$, then we find the lowest $j$ such that $x'_j > x_j$ and $B_1:= \{1,2,\dots,p'(x'_{j-1})\}$ is a ``decreasing" block.
            Moving along the list and repeating the process, we can then partition $\bm{z}$ into blocks $B_1,B_2,\dots$ which are alternately increasing and decreasing.
    
            We can say something about the index positions of the $x_i$ according to whether they are increasing or decreasing:
            \begin{align}
                x'_i > x_i  & \Rightarrow p'(x'_i) \leq p(x_i), \label{eqn:inc_block} \\
                x'_i \leq x_i & \Rightarrow p'(x'_i) \geq p(x_i). \label{eqn:dec_block}
            \end{align}
            To see this: suppose first that $p'(x'_i) < p(x_i)$, then there exists $y_j$ such that $y_j \geq x_i$ and $x'_i > y_j$ because $x_i$ has moved earlier in the list and thus displaced $y_j$ -- therefore $x'_i > x_i$.
            This proves~\eqref{eqn:dec_block} via its contrapositive.
            The same argument works for the opposite inequality~\eqref{eqn:inc_block}.
            If $p'(x'_i) = p(x_i)$, then either case is possible.
    
            Now our main claim is that the blocks $B_b$ have the following property:
            \begin{equation}
                \forall B_b, \quad p(x_i) \in B_b \Leftrightarrow p'(x'_i) \in B_b.
            \end{equation}
            In other words, the blocks simultaneously partition $\bm{z}$ and $\bm{z'}$ such that all pairs of $x_i$ and $x'_i$ are in the same blocks in their respective lists.
            We prove this claim by construction.
            \begin{itemize}
                \item If $B_1$ is increasing, then for all $x_i$ with $p(x_i) \in B_1$, we must have $p'(x'_i) \in B_1$ due to~\eqref{eqn:inc_block}.
                Also, whenever $p'(x'_i) \in B_1$, we must have $p(x_i) \in B_1$, otherwise $p'(x'_i)$ would be strictly less than $p(x_i)$ and by~\eqref{eqn:inc_block}, there would have been another increasing $x_i$ not included in the block. (It is also not possible to have some increasing $x_i$ in some later increasing block, for example $B_3$, since then there would be some $x_j$ in $B_2$ such that $x_j>x_i$ but $x'_i > x'_j$ -- and this is incompatible with the ordering $j<i$.)
                \item For the next block $B_2$, which is decreasing, by definition all $x'_i$ are included.
                We already know from the first step that there cannot be $x_i$ in $B_1$ such that $x'_i$ is in $B_2$.
                And by~\eqref{eqn:dec_block}, for any $x'_1$ in $B_2$, $p(x_i) \leq p'(x'_i)$, so $x_i$ cannot be outside of $B_2$.
                \item Iterate the procedure over all remaining blocks.
                \item If instead $B_1$ is decreasing, then the argument proceeds similarly.
            \end{itemize}

        \end{proof}
    \end{theorem}

    \textit{Proof of Theorem~\ref{thm:multi_strong_cat}}:
    First note that the sufficiency of the conditions~\eqref{eqn:strong_cat_approx_m} is straightforward: for each $\mu'_i \leq \mu_i$, we use the same construction as in Theorem~\ref{thm:no_catalyst}, while for any violation of that inequality, we simply do nothing to the corresponding mode.\\

    The proof of necessity for $A$ is an immediate application of Theorem~\ref{thm:approx_increases}, setting $\bm{x} = \bm{\alpha}_\mrm{S}$, $\bm{x'} = \bm{\alpha'}_\mrm{S}$, $\bm{y} = \bm{\alpha}_\mrm{C}$ and $\bm{\tilde{z}} = \bm{\tilde{\alpha}}_\mrm{SC}$.\\

    For $M$ we have more work to do because the eigenvalues must be divided into positive and negative parts.
    Unlike the non-catalytic case, it is possible that $\bm{\mu'}_\mrm{S}$ contains more positive or negative values than $\bm{\mu}_\mrm{S}$.
    If the number $n'_+$ of positive $\bm{\mu'}_\mrm{S}$ is greater than the number $n_+$ of positive $\bm{\mu}_\mrm{S}$, then we set $\bm{x} = (\bm{\mu}^+_\mrm{S},0,\dots)$, padding with zeroes to a total length of $n'_+$.
    We straightforwardly set $\bm{x'} = \bm{\mu'}^+_\mrm{S}$ and $\bm{y} = \bm{\mu}^+_\mrm{C}$.

    We also choose $\bm{\tilde{z}} = (\bm{\tilde{\mu}}^+_\mrm{SC},0,\dots)$, similarly padded to the same length as $\bm{z}$ -- and need to check that this satisfies the assumptions of Theorem~\ref{thm:approx_increases}.
    Since $M_\mrm{S} \oplus M_\mrm{C} \gto \tilde{M}_\mrm{SC}$, Theorem~\ref{thm:no_catalyst} ensures that the number of positive elements $\tilde{l}_+$ of $\bm{\tilde{\mu}}_\mrm{SC}$ is no more than the length $l_+$ of $\bm{z}$.
    Thus we have $\bm{\tilde{z}} \leq \bm{z}$.
    Moreover,
    \begin{align}
        \sum_{i=1}^{l_+} \abs{\tilde{z}_i - z'_i} & = \sum_{i\leq \tilde{l}_+} \abs{\tilde{z}_i - z'_i} + \sum_{i>\tilde{l}_+} \abs{\tilde{z}_i - z'_i} \nonumber \\
            & = \sum_{i \leq \tilde{l}_+} \abs{\tilde{\mu}^+_{\mrm{SC},i} - {\mu'}^+_{\mrm{SC},i}} + \sum_{i>\tilde{l}_+} {\mu'}^+_{\mrm{SC},i} - 0 \nonumber \\
            & \leq \sum_{i \leq \tilde{l}_+} \abs{\tilde{\mu}^+_{\mrm{SC},i} - {\mu'}^+_{\mrm{SC},i}} + \sum_{i>\tilde{l}_+} {\mu'}^+_{\mrm{SC},i} - \tilde{\mu}_{\mrm{SC},i} \nonumber \\
            & = \sum_{i=1}^{l_+} \abs{\mu'_{\mrm{SC},i} - \tilde{\mu}_{\mrm{SC},i}} =: \delta_+,
    \end{align}
    where the third line uses $\tilde{\mu}_{\mrm{SC},i} \leq 0$ for $i > \tilde{l}_+$.
    So Theorem~\ref{thm:approx_increases} can now be applied with the distance bound $\delta_+$.
    The same calculation can be done in exactly the same way for the negative values; summing the two results gives the claimed result since $\delta_+ + \delta_- \leq \delta$.

\section{Proof of Theorem~\ref{thm:M_weak_cat}} \label{app:M_weak_cat}
    We first prove the following useful characterisation of majorisation with an additive catalyst:
    \begin{lemma} \label{lem:catalyst_maj}
        For any $\bm{y}$, $(\bm{x'},\bm{y}) \prec_w (\bm{x},\bm{y}) \Leftrightarrow \bm{x'} \prec_w \bm{x}$ and $(\bm{x'},\bm{y}) \prec (\bm{x},\bm{y}) \Leftrightarrow \bm{x'} \prec \bm{x}$.
        \begin{proof}
            We use the following equivalent characterisation of majorisation~\cite[4.B.3]{MarshallBook}: $\bm{x} \prec_w \bm{y}$ if and only if
            \begin{equation}
                g_a(\bm{x}) \leq g_a(\bm{y}) \; \forall \, a \in \mathbb{R}, \text{ where } g_a(\bm{x}) := \sum_i \pos{x_i-a}.
            \end{equation}
            For $\bm{x} \prec \bm{y}$ we just have to add in the condition $\sum_i x_i = \sum_i y_i$.
            Using additivity of $g_a$,
            \begin{align}
                (\bm{x'},\bm{y}) \prec_w (\bm{x},\bm{y}) & \Leftrightarrow g_a(\bm{x'}) + g_a(\bm{y}) \leq g_a(\bm{x}) + g_a(\bm{y}) \; \forall a \nonumber \\
                & \Leftrightarrow g_a(\bm{x'}) \leq g_a(\bm{x}) \; \forall a \nonumber \\
                & \Leftrightarrow \bm{x'} \prec_w \bm{x}.
            \end{align}
            For $\bm{x} \prec \bm{x'}$, the sum condition is also clearly equivalent.
        \end{proof}
    \end{lemma}

    \textit{Proof of Theorem~\ref{thm:M_weak_cat}:}
    For part (1): The initial $M$ matrix is of the form $M = M_\mrm{S} \oplus M_\mrm{C}$, with eigenvalues $(\bm{\mu}_\mrm{S}, \bm{\mu}_\mrm{C})$.
    The final matrix is of the block form
    \begin{equation}
        M' = UMU^\dagger = \begin{pmatrix}
            M'_\mrm{S} & * \\ * & M_\mrm{C}
        \end{pmatrix}.
    \end{equation}
    Using local unitary rotations, we can diagonalise the principal blocks to obtain $(\bm{\mu'}_\mrm{S}, \bm{\mu'}_\mrm{C})$ on the diagonals of $M'$ (although the off-diagonal blocks need not vanish).
    The Schur-Horn theorem~\cite[Theorems 4.3.45, 4.3.48]{Horn1985Matrix} says that the eigenvalues of a matrix majorise its diagonals in any basis (and conversely that a basis can always be found giving any set of diagonals allowed by this condition).
    Hence $(\bm{\mu'}_\mrm{S},\bm{\mu}_\mrm{C}) \prec (\bm{\mu}_\mrm{S}, \bm{\mu}_\mrm{C})$ and Lemma~\ref{lem:catalyst_maj} gives the claimed condition as necessary.
    
    For the converse, we use the fact~\cite[2.B.1]{MarshallBook} that if $\bm{x} \prec \bm{y}$, then $\bm{x}$ can be obtained from $\bm{y}$ by a finite number (in fact, at most $n-1$) \emph{T-transforms}, namely partial swaps of pairs of modes -- represented by matrices of the form
    \begin{equation}
        T = tI + (1-t)Q,
    \end{equation}
    where $t \in [0,1]$ and $Q$ swaps two modes.
    In order to show that any T-transform can be performed, we focus on an arbitrary pair of modes, labelling the eigenvalues without loss of generality as $\mu_1 \geq \mu_2$.
    This is equivalent to proving that any $\mu'_1 \geq \mu'_2$ satisfying $\mu'_1 \leq \mu_1$ and $\mu'_1 + \mu'_2 = \mu_1 + \mu_2$ can be achieved with a single catalyst mode.
    This requires a unitary $U$ such that
    \begin{equation}
        U \begin{pmatrix}
            \mu_1 & 0 & 0 \\ 0 & \mu_2 & 0 \\ 0 & 0 & \mu_C
        \end{pmatrix} U^\dagger = 
        \begin{pmatrix}
            \mu'_1 & 0 & * \\ 0 & \mu'_2 & * \\ * & * & \mu_C
        \end{pmatrix}.
    \end{equation} 
    Choose any $\mu_C \in [\mu'_2, \mu'_1]$, which by assumption is also contained in $[\mu_2,\mu_1]$.
    The necessary and sufficient conditions for the existence of the eigenvalues $\mu'_1,\mu'_2$ of the two-dimensional upper-left block are~\cite[Theorem 4.3.21]{Horn1985Matrix}, bearing in mind the ordering $\mu_1 \geq \mu_C \geq \mu_2$,
    \begin{equation}
        \mu_1 \geq \mu'_1 \geq \mu_C  \geq \mu'_2 \geq \mu_2.
    \end{equation}
    Evidently these are always satisfied under the assumed conditions. \\

    Part (2) uses part (1) after extending the system to include the bath modes. This immediately gives the necessary condition
    \begin{equation} \label{eqn:mu_SB_maj}
        \bm{\mu'}_\mrm{SB} \prec \bm{\mu}_\mrm{SB} = (\bm{\mu}_\mrm{S}, \bm{0}_\mrm{B}).
    \end{equation}
    Since $\bm{\mu'}_\mrm{S}$ are eigenvalues of a principal submatrix of $M'_\mrm{SB}$, for any $k \leq n'_+$ we have~\cite[Corollary 4.3.34]{Horn1985Matrix} $\sum_{i=1}^k \mu'_{\mrm{S},i} \leq \sum_{i=1}^k \mu'_{\mrm{SB},i}$.
    Therefore
    \begin{align}
        \sum_{i=1}^k {\mu'}^+_{\mrm{S},i} & = \sum_{i=1}^k \mu'_{\mrm{S}, i} \nonumber \\
            & \leq \sum_{i=1}^k {\mu'}_{\mrm{SB},i} \nonumber \\
            & \leq \sum_{i=1}^k \mu_{\mrm{SB},i} \nonumber \\
            & \leq \sum_{i=1}^k \pos{\mu_{\mrm{SB},i}} \nonumber \\
            & = \sum_{i=1}^k \mu^+_{\mrm{S},i},
    \end{align}
    where the second inequality follows from \eqref{eqn:mu_SB_maj}.
    Hence $\bm{{\mu'}^+}_\mrm{S} \prec_w \bm{\mu^+}_\mrm{S}$.
    The corresponding necessary condition for the negative values follows by symmetry.

    For sufficiency, we use the following result~\cite[2.C.6.a]{MarshallBook} analogous to that used in part (1) but for weak majorisation: for $\bm{x},\bm{y}$ composed of non-negative elements, $\bm{x} \prec_w \bm{y}$ if and only if $\bm{x}$ can be derived from $\bm{y}$ by a finite number ($\leq n-1$) of T-transforms, followed by a finite number ($\leq n$) of L-transforms.
    The T-transforms were dealt with above; an L-transform can be performed simply by mixing a single mode at a beam splitter with a thermal mode as in Theorem~\ref{thm:no_catalyst}.
    The same construction works for the negative values independently.

\section{Proof of Theorem~\ref{thm:A_weak_cat}} \label{app:A_weak_cat}
    We prove necessity of the condition including a bath, since this is more general.
    As in Theorem~\ref{thm:M_weak_cat}, we have $A' = UAU^T$ for some unitary $U$ and the singular values of $A$ are $(\bm{\alpha}_\mrm{S},\bm{0}_\mrm{B},\bm{\alpha}_\mrm{C})$.
    The parameters $(\bm{\alpha'}_\mrm{SB},\bm{\alpha}_\mrm{C})$ are just diagonals of $A'$.
    A set of necessary conditions relating the diagonals $d_i$ of a complex symmetric matrix to its singular values $s_i$ are~\cite{Thompson1979Singular}
    \begin{equation}
            \sum_{i=1}^k \abs{d_i} \leq \sum_{i=1}^k s_i \quad \forall k.
    \end{equation}
    This implies $(\bm{\alpha'}_\mrm{SB}, \bm{\alpha}_C) \prec_w (\bm{\alpha}_\mrm{S},\bm{0}_\mrm{B},\bm{\alpha}_\mrm{C})$ and by Lemma~\ref{lem:catalyst_maj},
    \begin{equation}
        \bm{\alpha'}_\mrm{SB} \prec_w (\bm{\alpha}_\mrm{S}, \bm{0}_\mrm{B}).
    \end{equation}
    Therefore
    \begin{align}
        \sum_{i=1}^k \alpha'_{\mrm{S},i} & \leq \sum_{i=1}^k \alpha'_{\mrm{SB},i} \nonumber \\
            & \leq \sum_{i=1}^k \alpha_{\mrm{SB},i} \nonumber \\
            & = \sum_{i=1}^k \alpha_{\mrm{S},i}.
    \end{align}

    Sufficiency is proved without using a bath, using the same statement about weak majorisation used in Theorem~\ref{thm:M_weak_cat}.
    It is easy to see that T-transforms can be performed by just considering real matrix elements, so that $U$ is orthogonal and $A' = UAU^T$ is real and symmetric (since we can take $A$ as diagonal and containing its singular values without loss of generality).
    The same statement~\cite[Theorem 4.3.21]{Horn1985Matrix} also guarantees that the required sets of eigenvalues can be achieved with purely real matrix elements.
    L-transforms can be performed as before using a thermal bath mode.
    Alternatively, a catalyst mode can be used instead with the same unitary as in Eq.~\eqref{eqn:alpha_unitary} for the single-mode case.
    So with no bath, we need $2n-1$ catalyst modes, otherwise we need $n$ catalyst modes plus $n-1$ bath modes.

%%%%%%%%%%%%%
% \bibliographystyle{apsrev}
\bibliography{./bibliography}
%%%%%%%%%%%%%
\end{document}